\documentclass[runningheads]{llncs}        
\usepackage{amssymb}
\usepackage{amsmath}

\usepackage{amsthm}
\usepackage{proof}
\usepackage{stmaryrd}

\usepackage{booktabs}
\usepackage{dsfont}

\usepackage{bbm}
\usepackage{graphicx}
\usepackage{listings}
\usepackage{bussproofs}
\usepackage{tikz}
\usepackage{floatflt} 

\usepackage{ifthen}
\usepackage{xspace}
\usepackage{xcolor}
\usepackage{pdfpages}
\usepackage{todonotes}
\usepackage{afterpage}
\usepackage{paralist}
\usepackage{xspace}

\usepackage{pgf-umlsd}
\usetikzlibrary{arrows,shapes,trees,positioning,decorations,shapes}
\usetikzlibrary{decorations.markings,automata}

\DeclareOldFontCommand{\rm}{\normalfont\rmfamily}{\mathrm}
\DeclareOldFontCommand{\sf}{\normalfont\sffamily}{\mathsf}
\DeclareOldFontCommand{\tt}{\normalfont\ttfamily}{\mathtt}
\DeclareOldFontCommand{\bf}{\normalfont\bfseries}{\mathbf}
\DeclareOldFontCommand{\it}{\normalfont\itshape}{\mathit}
\DeclareOldFontCommand{\sl}{\normalfont\slshape}{\@nomath\sl}
\DeclareOldFontCommand{\sc}{\normalfont\scshape}{\@nomath\sc}
\DeclareRobustCommand*\cal{\@fontswitch\relax\mathcal}
\DeclareRobustCommand*\mit{\@fontswitch\relax\mathnormal}

\colorlet{keywordcolor}{blue!50!black}
\colorlet{commentcolor}{green!60!black}
\colorlet{typecolor}{violet}
\newcommand{\sourcefont}{\ttfamily\small}
\newcommand{\commentfont}{\slshape\rmfamily\color{commentcolor}}
\lstdefinelanguage{ABS}{
        keywords={physical,duration,diff,differential,do,assert,this,new,data,type,def,case,of,local,class,interface,
        extends,implements,if,then,else,await,get,Fut,return,skip,while,module,
        import,export,from,to,suspend,delta,adds,modifies,removes,original,productline,
        features,core,corefeatures,optionalfeatures,after,when,product,hasAttribute,
        hasMethod,hasField,hasInterface,uses,root,extension,group,allof,oneof,require,
        stateupdate,object,main,objectupdate,classupdate,fi,
        exclude,original,ifin,ifout,opt,null,%critical,port,rebind,duration,deadline,now,
        newgroup,data,thiscomp,in,joins,leaves,subtypeOf,wait,acquire,except,as,component,Pre,Abs
        },
        keywordstyle=\color{keywordcolor}\bfseries\sffamily,
        morekeywords=[2]{Unit, Int, Bool, Rat, List, Set, Pair, Fut, Maybe, String, Triple, Either, Map, Real},
        keywordstyle=[2]\color{typecolor},
        sensitive=true,
        comment=[l]{//},
        morecomment=[s]{/*}{*/},
        morestring=[b]"
}

\lstdefinelanguage[v9]{Java}[]{Java}{
        morekeywords={module,requires,provides,uses,with,to,exports}
}
\lstdefinelanguage[ContextJ]{Java}[]{Java}{
        morekeywords={layer,with,without,proceed,before,after}
}
\lstdefinelanguage[FOP]{Java}[]{Java}{
        morekeywords={refines,original,Super}
}
\lstdefinelanguage[JastAdd]{Java}[]{Java}{
        morekeywords={aspect,syn,inh,lazy}
}

\lstdefinestyle{code}{
        basicstyle=\sourcefont\upshape,
        keywordstyle=\color{keywordcolor}\bfseries\sffamily,
        commentstyle=\commentfont,
        columns=fullflexible,
        mathescape=true,
        escapechar={\#},
        keepspaces=true,
        showstringspaces=false,
        aboveskip=8pt, % default is \medskipamount
        numbers=left,
        stepnumber=1, 
        numberstyle=\ttfamily\scriptsize\color{gray},
        numbersep=4pt,
        xleftmargin=1.5em,
        xrightmargin=1.5em,
        framexleftmargin=1.2em,
        framexrightmargin=1em,
        framextopmargin=0.5ex,
        breaklines=true,
        breakindent=3pt,
}

\lstdefinestyle{abs}{
        style=code,
        language=ABS,
}
\lstdefinestyle{java}{
        style=code,
            language=Java
}
\lstdefinestyle{java9}{
        style=code,
            language=[v9]Java
}
\lstdefinestyle{aspectj}{
        style=code,
        language=[AspectJ]Java
}
\lstdefinestyle{jastadd}{
        style=code,
        language=[JastAdd]Java
}
\lstdefinestyle{contextj}{
        style=code,
        language=[ContextJ]Java
}
\lstdefinestyle{FOP}{
        style=code,
        language=[FOP]Java
}
\lstdefinestyle{scala}{
        style=code,
        language=Scala,
        morekeywords={self}
}

\newcommand{\code}[2][]{\lstinline[style=code,basicstyle=\ttfamily\upshape,#1]{#2}}

\newcommand{\abs}[2][]{\code[style=abs,#1]{#2}}

\lstnewenvironment{srccode}[1][]{
        \minipage{\linewidth}
        \lstset{style=code,
        backgroundcolor=\color{white},
        rulecolor=\color{gray!50},
        frame=tblr,
        captionpos=b,
        #1}
}{\endminipage}

\lstnewenvironment{abscode}[1][]{
        \minipage{1\linewidth}
        \lstset{style=abs,
        backgroundcolor=\color{white},
        rulecolor=\color{gray!50},
        frame=tblr,
        captionpos=b,
        #1}
}{\endminipage}

\lstnewenvironment{javacode}[1][]{
        \minipage{\linewidth}
        \lstset{style=java,
        backgroundcolor=\color{gray!5},
        rulecolor=\color{gray!50},
        frame=tblr,
        captionpos=b,
        #1}
}{\endminipage}

\lstnewenvironment{jastaddcode}[1][]{
        \minipage{\linewidth}
        \lstset{style=jastadd,
        backgroundcolor=\color{gray!5},
        rulecolor=\color{gray!50},
        frame=tblr,
        captionpos=b,
        #1}
}{\endminipage}

 \makeatletter
\newcommand\BeraMonottfamily{%
  \def\fvm@Scale{0.85}% scales the font down
  \fontfamily{fvm}\selectfont% selects the Bera Mono font
}
\makeatother

\lstdefinelanguage{KeYmaeraX}{%
  keywords={if,then,else,R,B,HP,Functions,ProgramVariables,Problem,End,Definitions,ArchiveEntry,Tactic,SharedDefinitions},%
  sensitive=true,
  morecomment=[s]{/*}{*/},
  deletestring=[d]',
  showstringspaces=false,
  commentstyle=\color{green},
  mathescape,
  escapeinside={/*@}{@*/}}[keywords]

\lstdefinelanguage{Bellerophon}{%
  language={},
  keywords={'R,'L,'_},%
  otherkeywords={;,<,|},
  sensitive=true,
  morecomment=[l]{//},
  morecomment=[s]{/*}{*/},
  morestring=[b]",
  deletestring=[d]',
  morestring=[d]`,
  showstringspaces=false,
  commentstyle=\fontseries{lc}\color{green}}[keywords]
  
\newcommand{\keycode}[1]{
    { \lstset{language=KeYmaeraX}
    \begin{lstlisting}
     #1
    \end{lstlisting}
    }
}

\newcommand{\COMMENT}[1]{}

\DeclareMathOperator*{\argmin}{\mathbf{argmin}}

\let\temp\phi
\let\phi\varphi
\let\varphi\temp

\makeatletter
\newcommand{\xRightarrow}[2][]{\ext@arrow 0359\Rightarrowfill@{#1}{#2}}

\newcommand{\cased}[1]{\ensuremath{
\left\{\begin{array}{ll}
#1
\end{array}\right.
}\xspace}

\newcommand{\xabs}[1]{\text{\abs{#1}}}
\newcommand{\kyx}[1]{\mathbf{#1}}
\newcommand{\sem}[1]{\ensuremath{ \llbracket #1 \rrbracket \xspace}}
\newcommand{\sep}{  \ | \ }
\newcommand{\trans}{\ensuremath{\mathsf{trans}}\xspace}
\newcommand{\transx}[1]{\ensuremath{\mathsf{trans}(\xabs{#1})}\xspace}
\newcommand{\many}[1]{\overline{#1}}

\newcommand{\ddl}{\ensuremath{d\mathcal{L}}\xspace}
\newcommand{\dL}{\ddl}
\newcommand{\QdL}{Q\ddl}
\newcommand{\KeYmaeraX}{KeYmaera~X\xspace}

\newcommand*{\ptest}[1]{\ensuremath{?#1}}
\newcommand*{\prepeat}[2][*]{\ensuremath{{#2}^{#1}}}
\newcommand*{\pumod}[2]
	{#1\hspace{-0.09em}\mathrel{{:}{=}}\hspace{-0.09em}#2}
\newcommand*{\prandom}[1]{\pumod{#1}{\ast}}
\newcommand*{\pchoice}[2]{\ensuremath{{#1}\cup{#2}}}
\newcommand*{\pevolvein}[2]{\ensuremath{\{#1 \,\&\, #2\}}}
\newcommand*{\D}[1]{#1'}
\newcommand*{\limply}{\ensuremath{\rightarrow}}
\newcommand*{\dbox}[2]{\ensuremath{[#1]#2}}

\def\myttsize{\fontsize{8}{9}}
\newcommand{\key}[1]{\mbox{\ttfamily\bfseries #1}}
\newcommand{\append}{\circ}
\newcommand{\geval}[3]{\mbox{$[\![#1]\!]_{#2}^{#3}$}}
\newcommand{\eval}[2]{\mbox{$[\![#1]\!]_{#2}$}}

\newcommand{\arulename}[1]{\mbox{\footnotesize\scshape #1}}
\newcommand{\none}{\varepsilon}
\newcommand{\nt}[1]{\mathit{#1}}
\newcommand{\asep}{\:|\:}
\newcommand{\TYPE}[1]{\mbox{\sffamily #1}}
\newcommand{\option}[1]{[#1]}

\newcommand{\ntyperule}[3]{ 
  \begin{array}{c} 
    \textsc{\scriptsize ({#1})} \\ 
    #2 \\ 
    \hline
    #3 
  \end{array}} 

\newcommand{\nredrule}[2]{ 
  \begin{array}{c} 
    \textsc{\scriptsize ({#1})} \\ 
    #2  
  \end{array}} 

\title{Modeling and Verifying Cyber-Physical~Systems with Hybrid~Active~Objects\thanks{This work is partially supported by the \texttt{FormbaR} project, part of AG Signalling/DB RailLab in the Innovation Alliance of Deutsche Bahn AG and TU Darmstadt. This material is based upon work supported by AFOSR grant FA9550-16-1-0288.}}
\author{
     Eduard~Kamburjan\inst{1} 
\and Stefan~Mitsch\inst{2}
\and Martina~Kettenbach\inst{1} 
\and Reiner~H{\"a}hnle\inst{1} 
}
\institute{
Department of Computer Science, Technische Universit\"at Darmstadt, Germany \\
  \email{\{haehnle,kamburjan\}@cs.tu-darmstadt.de}\quad \email{martina.kettenbach@gmail.com}
  \and
Computer Science Department, Carnegie Mellon University, USA \\
  \email{smitsch@cs.cmu.edu}
}
\begin{document}
\maketitle
\begin{abstract}
  Formal modeling of cyber-physical systems (CPS) is hard, because
  they pose the double challenge of combined discrete-continous
  dynamics and concurrent behavior. Existing formal specification and
  verification languages for CPS are designed on top of their
  underlying proof search technology. They lack high-level structuring
  elements. In addition, they are not or not efficiently
  executable. This makes formal CPS models hard to understand and to
  validate, hence impairs their usability. Instead, we suggest to
  model CPS in an Active Objects (AO) language designed for concise,
  intuitive modeling of concurrent systems. To this end, we extend the
  AO language ABS and its runtime environment with Hybrid Active
  Objects (HAO). CPS models and requirements formalized in HAO must
  follow certain communication patterns that permit automatic
  translation into differential dynamic logic, a sequential hybrid
  program logic. Verification is achieved by discharging the resulting
  formulas with the theorem prover \KeYmaeraX. We demonstrate the
  practicality of our approach with case studies.
  %
  \iffalse
  %
  This paper presents \emph{Hybrid Active Objects} (HAO), a formalism
  to model, simulate and verify distributed systems with mixed
  discrete and continuous behavior in an object-oriented way.  Active
  Objects are strongly encapsulated actors with cooperative scheduling
  and Hybrid Active Objects use physical fields to describe continuous
  state change with ODEs.  As a program-based formalism, HAOs are more
  natural than established formalisms for cyber-physical systems and,
  thus, lower the cognitive burden needed for formal modeling.  We
  implement Hybrid Active Objects in the HABS language, an extension
  of the ABS language.  HABS allows to simulate and visualize systems
  consisting of Hybrid Active Objects and can be verified by a
  translation into differential dynamic logic, which is supported by
  the KeYmaera X tool.  We present a case study with a simple
  distributed water tank that showcases the tool-supported simulation
  and verification.
  %
  \fi
\end{abstract}

\section{Introduction}\label{sec:intro}
Formal modeling of cyber-physical systems (CPS) poses a double
challenge: first, their \emph{hybrid} nature, with both continuous
physical dynamics and complex computations in discrete time
steps. Second, their \emph{concurrent} nature: distributed, active
components (sensors, actuators, controllers) execute simultaneously
and communicate asynchronously. For this reason, it is particularly
difficult to get models of CPS right.  But to be useful in practice, a
formal modeling language must support \emph{validation} \cite{DBLP:journals/fmsd/MitschP16}, i.e.\ to
ensure that the model correctly captures a system and its
requirements. Existing modeling languages for CPS, however, are
designed for \emph{verification}: to prove that a formal model
satisfies formal properties. Examples include hybrid
automata~\cite{Rajeev93}, hybrid process algebra~\cite{Cuijpers05},
and logics for hybrid programs~\cite{Platzer10}. Models written in
these languages are hard to read, because they are expressed in terms
of the underlying verification technology: automata, algebras,
formulas. They lack high-level structuring elements such as types,
scopes, methods, complex commands, futures, etc. In addition, these
languages are not or not efficiently executable, a crucial aspect of
validation.

\iffalse
%
Distributed and component-based Cyber-physical systems (CPS) exhibit a
hybrid nature, having both continuous physical dynamics and complex
computations in discrete time steps, which poses a challenge to formal
modeling, simulation, and verification.

State-of-the-art modeling languages for formal verification of CPS are
low-level languages, such as hybrid automata~\cite{Rajeev93}, hybrid
process algebras~\cite{Cuijpers05} or logics for hybrid
programs~\cite{Platzer10}.  They do not provide extensive support in
terms of usability known from modern programming languages, and are
not integrated into a modeling, simulation, and verification workflow
with short feedback cycles.  They are unsuitable for explorative
modeling or formal prototyping~\cite{KamburjanH17a}.

% TOXDO relevant why?  Most approaches support some form of
% compositional \emph{modeling}~\cite{Lynch03,Schiffelers04}, but only
% few, (e.g., \cite{DBLP:journals/sttt/MullerMRSP18}) allow
% compositional \emph{verification}.  Compositional verification,
% however, is critical for serious verification attempts, especially
% for distributed systems.
%
\fi

We make a proposal to address the ``usability gap'' in modeling CPS
while retaining the capability to formally verify properties of
models. Our starting point is the Active Objects
\cite{BoerSHHRDJSKFY17} language ABS (for: Abstract Behavioral
Specification)~\cite{JohnsenHSSS10}. It is an efficiently executable
language, designed to model asynchronous, concurrent systems. ABS was
used to model complex, real-world systems for cloud processing
\cite{AlbertBHJSTW14}, virtualized services~\cite{JohnsenPT17}, data
processing \cite{LinYJL16}, and railway operations~\cite{KHS18}.  It
is designed for verifiability and has associated analysis and
verification tools \cite{WongAMPSS12}, but it lacks the capability to
model hybrid systems.
%
\iffalse
%
For software-based systems, Active Objects~\cite{BoerSHHRDJSKFY17}
have been developed as a concurrency model that integrates modeling,
simulation and verification.  The Abstract Behavioral Specification
(ABS)~\cite{JohnsenHSSS10} language implements Active Objects and is
developed with a focus on usability.  Furthermore, ABS has a number of
static analyses~\cite{AlbertAFGGMPR14} and a rich program
logic~\cite{DinO15}, suitable for non-trivial verification.  It has
been successfully applied in the modeling (and partially analysis) of
frameworks for large data~\cite{LinLYJ18,LinYJL16} and virtualized
services~\cite{AlbertBHJSTW14,JohnsenPT17}.  A recent project on
verification of railway operations~\cite{KamburjanH17} has, however,
shown its limits for systems with continuous dynamics, which must be
discretized by hand and on which the available tools do not perform
well.
%
\fi
%
That is offered by our suggested \emph{Hybrid ABS} (HABS) language
extension, generalizing Active Objects to \emph{Hybrid Active Objects}
(HAO): Active Objects extended with continuous dynamics. However, we
do not endeavor to extend the existing verification system of ABS
\cite{DinBH15} to handle HAO. For once, this is a highly complex task;
second, an expressive hybrid verification logic is \emph{available} as
differential dynamic logic (\ddl)~\cite{DBLP:conf/lics/Platzer12b,DBLP:journals/jar/Platzer17,Platzer18},
implemented in the \KeYmaeraX system~\cite{FultonMQVP15}.

\begin{figure}[thb]
\centering\includegraphics[scale=0.5]{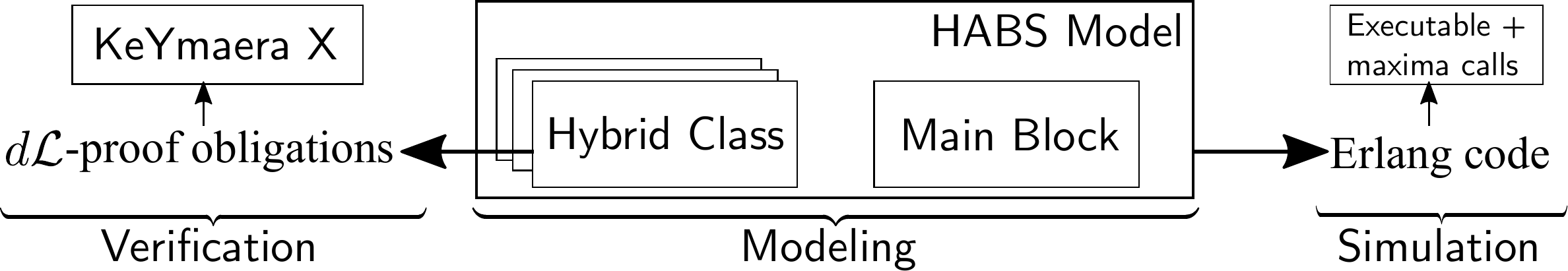}
\caption{Overview over HABS.}
\label{fig:over}
\end{figure}

Our approach is based on \emph{translation} from HABS to \ddl: Hybrid
models are developed and debugged in HABS---whose runtime environment
is extended to handle HAO---and then formally verified against a
safety property in \KeYmaeraX.  Fig.~\ref{fig:over} illustrates this:
Verification ensures that all HAO created to satisfy a given
precondition fulfill a given object invariant.  An optional main block
can be used to initialize a system model. This allows to verify
concrete scenarios.
Simulation of HABS models features a visualization component that
produces graphs showing how the value of object fields changes over
time.

\KeYmaeraX\ does not yet support \QdL~\cite{DBLP:journals/lmcs/Platzer12b} for distributed hybrid systems and its component-based techniques \cite{DBLP:journals/sttt/MullerMRSP18} are subject to strict interaction requirements. We solve this
by identifying an interaction pattern for communication that we impose
on HABS.  It is general enough to permit intuitive and concise
modeling of relevant case studies. The advantage of the pattern-based
approach is that one can decompose the HABS verification problem into
a set of independent \emph{sequential} \ddl\ problems for each hybrid
class. In this sense, verification is compositional.
%
% We give a modeling pattern for interactions that allows for multiple
% HAOs to communicate while classes are still verified in isolation.
% With the main block, the program may also be simulated. A HABS
% program is compiled into Erlang (with calls to Maxima~\cite{maxim}
% to analyze continuous behavior) and can be executed.

% By combining compositional verification, simulation and
% object-oriented modeling, HAOs are a natural concurrency model for
% distributed CPS and HABS offers tool-support for formal models.

The paper is structured as follows.  Sect.~\ref{sec:prelim} provides
background on \ddl. Sect.~\ref{sec:habs} gives syntax and semantics of
HABS, illustrated with water tank %by examples of water tank
models. Sect.~\ref{sec:translate} describes verification,
Sect.~\ref{sec:impl} compilation and
simulation. Sect.~\ref{sec:conclusion} concludes.

% For a large fragment, we are able to enable compositional
% verification by a translation from HAOs into components in
% differential dynamic logic~\cite{MullerMRSP16,Platzer10}.  We report
% on a prototypical implementation, Hybrid ABS (HABS), which extends
% Timed ABS~\cite{BjorkBJST13} and uses \KeYmaeraX~\cite{FultonMQVP15}
% for verification.  HABS also allows simulation and we illustrate
% HAOs with a case study for component-based modeling and verification
% of a water tank controller.

%%% Local Variables:
%%% mode: latex
%%% TeX-master: "paper"
%%% End:

\section{Background: Differential Dynamic Logic}\label{sec:prelim}
We briefly review \emph{differential dynamic logic} (\ddl)
\cite{DBLP:journals/jar/Platzer17,Platzer18} as implemented in the
interactive theorem prover \KeYmaeraX\ \cite{FultonMQVP15}.
Differential dynamic logic expresses the combined discrete and
continuous dynamics of hybrid systems in a sequential imperative
programming language called \emph{hybrid programs}.  Its syntax and
informal semantics are in Table~\ref{tab:hybrid-programs}.

\begin{table}[htb]
\centering
\caption{Hybrid programs in \ddl}
\label{tab:hybrid-programs}
\begin{tabular}{ll}
Program & Informal semantics   \\\hline
$\ptest{\phi}$         & Test whether formula $\phi$ is true, abort if false\\
$\pumod{x}{\theta}$    & Assign value of term $\theta$ to variable $x$\\
$\prandom{x}$          & Assign any (real) value to variable $x$\\
$\pevolvein{\D{x}=\theta}{H}$& Evolve ODE $\D{x} = \theta$ for any duration $t{\geq}0$ \\
& with evolution domain constraint $H$ true throughout\\
$\alpha;\beta$         & Run $\alpha$ followed by $\beta$ on resulting state(s)\\
$\alpha\cup\beta$      & Run either $\alpha$ or $\beta$ non-deterministically\\
$\prepeat{\alpha}$     & Repeat $\alpha$ $n$ times, for any $n\in\mathbb{N}$
\end{tabular}
\end{table}

Hybrid programs provide the usual discrete statements assignment
($\pumod{x}{\theta}$), non-deterministic assignment ($\prandom{x}$),
test ($\ptest{\phi}$), non-deterministic choice ($\alpha\cup\beta$),
sequential composition ($\alpha ; \beta$), and non-deterministic
repetition ($\prepeat{\alpha}$).  A typical modeling pattern combines
non-deterministic assignment and test (e.g., ``$\prandom{x};\ptest{H}$'')
to choose any value subject to a Boolean constraint
$H$.  Standard control structures are expressible, for example:
\begin{inparaenum}[(i)]
\item
  $\kyx{if}~ H ~\kyx{then}~ \alpha ~\kyx{else}~ \beta \equiv
  \pchoice{(\ptest{H};\alpha)}{(\ptest{\lnot H};\beta)}$,
\item
  $\kyx{if}~ H ~\kyx{then}~ \alpha \equiv
  \pchoice{(\ptest{H};\alpha)}{(\ptest{\lnot H})}$,
\item
  $\kyx{while}~(H) ~\alpha \equiv
  \prepeat{(\ptest{H};\alpha)};\ptest{\lnot H}$.
\end{inparaenum}

For continuous dynamics, the notation $\pevolvein{\D{x} = \theta}{H}$
represents an ordinary differential equation (ODE) system (derivative
$\D{x}$ in time) of the form $\D{x}_1 = \theta_1$, \dots,
$\D{x}_n = \theta_n$.  Any behavior described by the ODE stays inside
the evolution domain $H$, i.e.\ the ODE is followed for a
non-deterministic, non-negative period of time, but stops before $H$
becomes false.  For example, a basic model of the water level $x$ in a
tank draining with flow $-f$ is given by the ODE
$\pevolvein{\D{x}=-f}{x \geq 0}$, where the evolution domain
constraint $x \geq 0$ means the tank will not drain to negative water
levels.
% \ektodo{Maybe another example so the reader doesn't confuse it with
% the following water tanks?}
With a careful modeling pattern, continuous behavior can be governed by $H$ so that one can react to
events, but it is not otherwise restricted or influenced: The pattern
$\pchoice{\pevolvein{\D{x} = \theta}{H}}{\pevolvein{\D{x} =
    \theta}{\widetilde{H}}}$ permits control intervention to achieve
different behavior triggered by an event.  $\widetilde{H}$ is the weak
complement of~$H$: they share exactly their boundary from which both
behaviors are possible. For example, $H\equiv x\leq 0$,
$\widetilde{H}\equiv x\geq 0$.

The \dL-formulas $\phi,\,\psi$ relevant for this paper are
propositional logic operators $\phi\land\psi$, $\phi\lor\psi$,
$\phi\limply\psi$, $\neg\phi$ and
%$\forall{x}{~\phi}$, and $\exists{x}{~\phi}$ 
comparison expressions $\theta\sim\eta$, where
$\sim\mathop{\in}\{<,\,\leq,\,=,\,\neq,\,\geq,\,>\}$ and
$\theta,\,\eta$ are real-valued terms over $\{+,\,-,\,\cdot\,,/\}$. In
addition, there is the \dL\ modal operator $\dbox{\alpha}{\phi}$.
\iffalse
%
\begin{equation*}
  \phi,\psi ::= \phi\land\psi\ |\ \phi\lor\psi\ |\ \phi\limply\psi\ |\ \neg\phi\ |\ \theta \sim \eta
  % |\ \forall{x}{~\phi}\ |\ \exists{x}{~\phi}\ 
  |\ \dbox{\alpha}{\phi}\
\end{equation*}
%
\fi
%
The \dL-formula $\dbox{\alpha}{\phi}$ is true iff $\phi$ holds in all
states reachable by program $\alpha$.
The formal semantics of $\dL$
\cite{DBLP:journals/jar/Platzer17,Platzer18} is a Kripke semantics in
which the states of the Kripke model are the states of the hybrid
system.
% States are maps $\nu:\allvars\to\reals$ assigning a real value
% $\nu(x)$ to each variable $x\in\allvars$.
The semantics of a hybrid program $\alpha$ is a relation
$\llbracket\alpha\rrbracket$ between its initial and final states.
Specifically, $\nu\models\dbox{\alpha}\phi$ iff $\omega\models\phi$
for all states $(\nu,\omega) \in \llbracket\alpha\rrbracket$, so all
runs of $\alpha$ from $\nu$ are safe relative to $\phi$.

%%% Local Variables:
%%% mode: latex
%%% TeX-master: "paper"
%%% End:

\section{Hybrid Active Objects}\label{sec:habs}

% \subsection{Hybrid Active Objects}
% \label{sec:hybr-active-objects}

Active Objects~\cite{BoerSHHRDJSKFY17} are strongly encapsulated
objects that realize actor-based concurrency~\cite{actor} with
futures~\cite{fut} and cooperative scheduling:
Active Objects communicate via asynchronous method calls. On the
caller side of a method invocation each time a future is generated as
a handle to the call's result once it is available.  The caller may
synchronize on that future, i.e.\ suspend and wait until it is
resolved.
% Processes running on an Active Object use cooperative scheduling:
At most one process is running on an Active Object at any time. It
only suspends when it encounters the synchronization statement
\abs{await} on an unresolved future or a false Boolean condition.
Once the guard becomes true, the process may be re-scheduled. All
fields are strictly object-private.

Hybrid Active Objects (HAO) extend Active Objects with continuous
behavior, expressed by ODE over ``physical'' fields.  These change
their value following their ODE whenever time passes, even when no
process is active.
% no process explicitly changes their value or no process is active.

\begin{figure}[htb]
\begin{minipage}{0.5\textwidth}
\begin{abscode}[basicstyle=\ttfamily\scriptsize\upshape,numbers=none]
/*@ requires 4 < inVal & inVal < 9 @*/
class CSingleTank(Real inVal) {
/*@ invariant 3 <= level & level <= 10 @*/
 physical {
  Real level = inVal : level' = drain;
  Real drain = -1/2  : drain' = 0;
 }
 Unit run() { this!ctrl(); }
 Unit ctrl() { 
  await diff (level <= 3 & drain <= 0) 
        | (level >= 10 & drain >= 0);
  if (level <= 3) drain =  1/2;
  else            drain = -1/2;
  this.ctrl();
 }
}
\end{abscode}
\end{minipage}
\begin{minipage}{0.5\textwidth}
\includegraphics[scale=0.14]{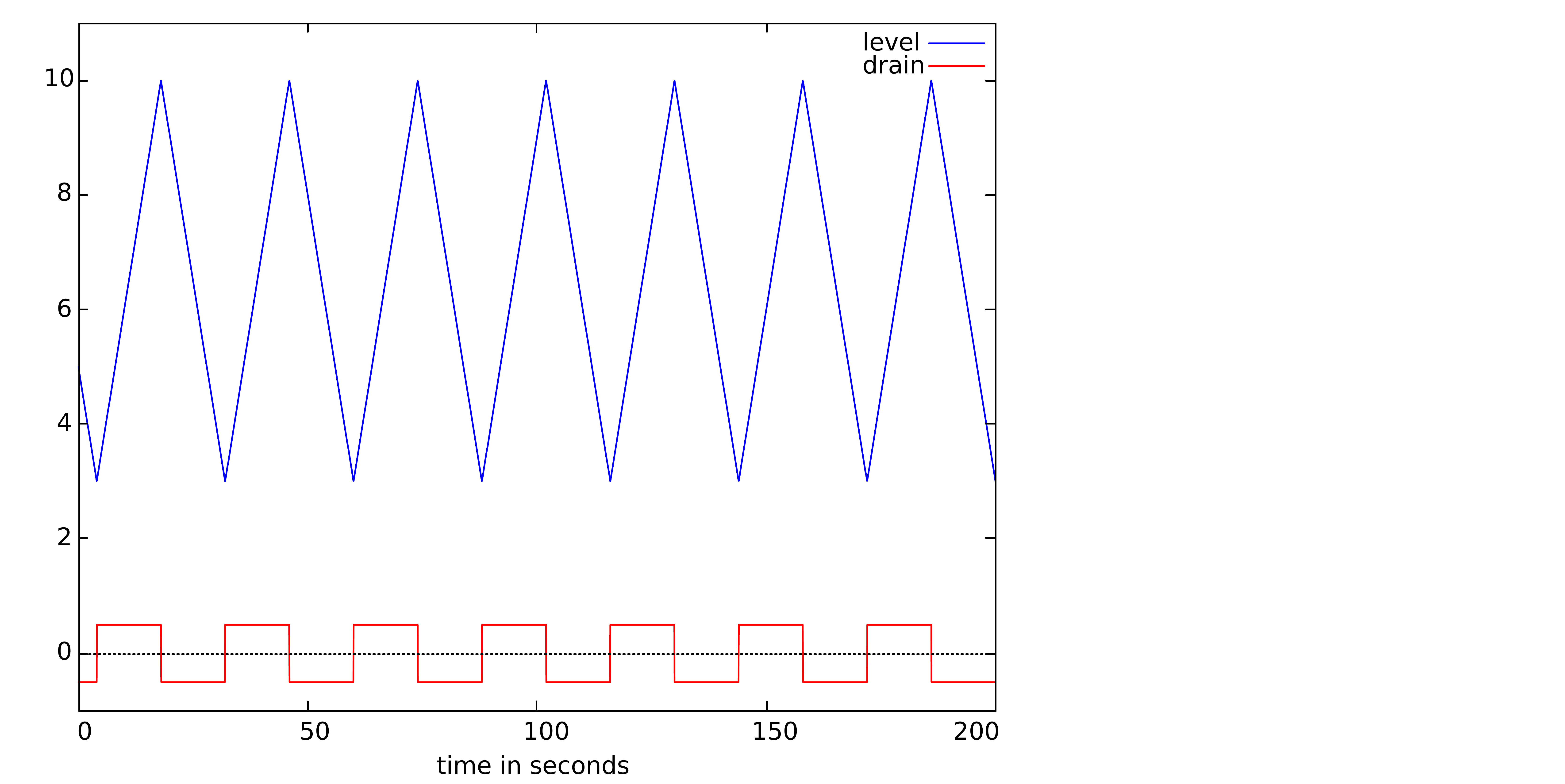}
\end{minipage}
\caption{A water tank as a Hybrid Active Object and a simulation output.}
\label{fig:tank1}
\end{figure}

\begin{example}\label{ex:tank1}
  Fig.~\ref{fig:tank1} shows on the left an HAO model of a water tank
  that either fills with $\frac{1}{2}l/sec$ or is drained with the same
  speed. Method \abs{ctrl()} realizes a control loop that sets the
  \abs{drain} field to switch between those states so that the level
  stays between 3$l$ and 10$l$.  Fields \abs{level}, \abs{drain} are
  physical. Their initial value and governing ODE is declared in the
  \abs{physical} section.  The controller suspends until %
  \abs{level <= 3 | level >= 10} holds, i.e.\ until the water level
  reaches the upper or lower limit.  Depending on which is the case,
  it changes the state and calls itself recursively.

  The JML style comments provide an assumption on the initial state
  and the safety condition that in this case can be proven: if the
  initial level is between 4$l$ and 9$l$, then it always stays between
  3$l$ and 10$l$. Before one attempts to formally verify this property
  one typically wants to run some tests to see whether it behaves as
  intended. Our HABS implementation allows to simulate and visualize an
  HAO model. For example, the graph in Fig.~\ref{fig:tank1} on the
  right shows the behavior of a \abs{CSingleTank} object instantiated
  with $\xabs{inVal}=5$. 
  In Sec.~\ref{sec:translate} we show how the class is translated
  into \ddl and prove the safety condition in \KeYmaeraX\ for
  \emph{any} object created with a parameter that satisfies the
  precondition.
\end{example}

\subsection{Hybrid ABS}

We formally define the Hybrid Abstract Behavioral Specification (HABS)
language based on ABS~\cite{JohnsenHSSS10}.  The syntax is in
Fig.~\ref{fig:syntax}. With \abs{e} we denote standard expressions
over fields \abs{f}, variables \abs{v} and operators \abs{|},
\abs{\&}, \abs{>=}, \abs{<=}, \abs{+}, \abs{-}, \abs{*}, \abs{/}.
Types \abs{T} are all interface names and \abs{Real}, \abs{Unit}. HABS
uses futures implicitly, no explicit future types need to be declared.

A program contains a main method, interfaces, classes. Interfaces are
standard, the main method contains a list of object creations.
Classes can have parameters, these are fields being initialized during
object creation. Classes have an optional physical block that declares
ODEs over physical fields. Classes can have further field declarations
and declare three different kinds of methods:
\begin{itemize}
\item \emph{Controllers}, starting with an \abs{await} statement and
  ending with a recursive call.
\item \emph{Ports}, used to communicate with other objects.
\item A mandatory \abs{Unit run()} method that only calls
  controllers. It is started upon object creation.
\end{itemize}

\begin{figure}[t!bh]
\centering
\scalebox{0.85}{
\begin{minipage}{\textwidth}
\begin{align*}
\mathsf{Prgm} ::=~& \many{\mathsf{ID}}~\many{\mathsf{CD}}~\mathsf{Main} \qquad \mathsf{ID} ::=~ \xabs{interface I}\left[\xabs{extends}~\many{\xabs{I}}\right]\{ \many{\mathsf{MS}}\} && \text{\small Program, Interfaces}\\
\mathsf{CD} ::=~& \xabs{class C}\left[\xabs{implements}~\many{\xabs{I}}\right]\left[\xabs{(}\many{\xabs{T v}}\xabs{)}\right]\{\mathsf{Phys}?~\many{\mathsf{FD}}~\many{\mathsf{Ctrl}}~\many{\mathsf{Port}}~\mathsf{Run}\} &&\text{\small Classes}\\
\mathsf{Main}::=~&\{\many{\xabs{I v = new C(}\many{\xabs{e}}\xabs{)}}\}\quad
\quad \mathsf{FD} ::=~ \xabs{T v = e}  
&& \text{\small Main, Fields}\\
\mathsf{Phys} ::=~& \xabs{physical}~\{\many{\mathsf{DED}}\} \qquad
\mathsf{DED} ::=~ \xabs{Real v = e}~:~\mathsf{de} \qquad \mathsf{de} ::=~ \xabs{e} \sep \xabs{e'} &&\text{\small Physical Block}\\
\midrule
\mathsf{MS} ::=~& \xabs{T m(}\many{T~\xabs{v}}\xabs{)} \qquad
\mathsf{Port} ::=~\mathsf{MS}~\{\xabs{return f;}\} \sep \mathsf{MS}~\{\xabs{f = v;}\} 
&&\text{\small  Signature, Ports}\\
\mathsf{Ctrl} ::=~&\mathsf{MS}~\{\mathsf{sa}\xabs{;}~\mathsf{s}\xabs{; this.m();}\}\qquad
\mathsf{Run}::=~\xabs{Unit run()}~\{\mathsf{sc}\} 
&&\text{\small Controllers, Run}\\
\midrule
\mathsf{sa} ::=~&\xabs{await duration(e,e)} \sep \xabs{await diff e}  \qquad
\mathsf{sc} ::=~\xabs{e!m(}\many{\xabs{e}}\xabs{)} \sep \mathsf{sc}\xabs{;}\mathsf{sc}&&\text{\small Suspension, Calls}\\
%\mathsf{se} ::=~[T]~\xabs{v = e.m()} \sep \xabs{s} \sep \mathsf{se}\xabs{;}\mathsf{se} 
%&&\text{\small Calling Statements}\\
\mathsf{s} ::=~&[\xabs{T}]~\xabs{v = e}\sep\xabs{v = e} 
\sep \xabs{f = e.m()} \sep \xabs{e!m(f)} 
&&\text{\small Atomic Statements}\\
\sep~&\xabs{while (e)}~\{\mathsf{s}\}
\sep\xabs{if (e)}~\{\mathsf{s}\}~[\xabs{else}~\{\mathsf{s}\}] 
\sep \mathsf{s}\xabs{;}\mathsf{s}
&&\text{\small Control Flow}
\end{align*}
\end{minipage}
}
\caption{Syntax of HABS}
\label{fig:syntax}
\end{figure}

% A controller may read from port methods of other directly after its
% \abs{await} suspension statement and write to port methods of other
% methods directly before its recurisve call. Inbetween, no calls and
% suspensions are allowed.
Controller methods realize control loops. They may call port methods
of other objects in each round after the initial \abs{await} suspension
statement.  Port methods neither suspend nor call other methods. The
first rule in the grammar for $\mathsf{Port}$ declares
\emph{out-port}, the second \emph{in-port} methods.  An out-port
returns a field, an in-port copies its parameter into a field.

There are two kinds of guards: \abs{diff e} suspends until the
expression \abs{e} evaluates to true, \abs{duration(e1,e2)} suspends
\COMMENT{for at least} between \abs{e1} and \abs{e2} time units.  The call
\abs{e.m} is shorthand for an asynchronous call directly followed by
synchronization on its future. The call \abs{e!m} is asynchronous and
\abs{this.m} is a standard internal synchronous call.  Each HABS
program is well-typed according to~\cite{BjorkBJST13,JohnsenHSSS10}
plus: % on controller methods:
\begin{inparaenum}[(i)]
\item the method calls at the end of controllers implement direct
  recursion;
\item the structure of controllers is to read from out-ports of other
  objects into local fields, followed by computation, followed by
  writing into in-ports of other objects.
\end{inparaenum}
By convention we let all controller method names start with
\abs{ctrl}, all in-ports with \abs[deletekeywords={in}]{in} and all
out-ports with \abs{out}.

\begin{figure}[thb]
\begin{minipage}{0.53\textwidth}
\begin{abscode}[basicstyle=\ttfamily\scriptsize\upshape]
interface Tank {
/*  requires -1/2 <= newD <= 1/2; */
 Unit inDrain(Real newD);
/* ensures 3 <= outLevel <= 10; */
 Real outLevel();
}
class Tank(Real inVal) implements Tank {
 physical {
   Real level = inVal : level' = drain;
   Real drain = -1/2 : drain' = 0;
 }
 Unit run() { }
/* requires newD > 0 -> level <= 9.5  */
/* requires newD < 0 -> level >= 3.5 */
/* timed_requires inDrain < 1 */
 Unit inDrain(Real newD) { drain = newD; }
 Real outLevel() { return level; }
}
\end{abscode}
\end{minipage}
\begin{minipage}{0.47\textwidth}
\begin{abscode}[basicstyle=\ttfamily\scriptsize\upshape,firstnumber=19]
/* requires 0 < tick < 1 */
class FlowCtrl(Tank t, Real tick){

/* invariant drain > 0 -> level <= 9.5 
            & drain < 0 -> level >= 3.5 */
  Real drain = 1/2; 
  Real level = 0;

  Unit run() { this!ctrlFlow(); }
  Unit ctrlFlow() {
    await duration(tick,tick);
    level = t.outLevel();
    if (level <= 3.5)  drain =  1/2; 
    if (level >= 9.5) drain = -1/2;
    t!inDrain(drain);  
    this.ctrlFlow();
  }
}
\end{abscode}
\end{minipage}
\caption{A water tank modeled as two Hybrid Active Objects. Invariant
  and precondition of \abs{Tank} are as in Fig.~\ref{fig:tank1}.}
\label{fig:tank2}
\end{figure}

\begin{example}\label{ex:tank2}
  Fig.~\ref{fig:tank2} shows a water tank realized by two objects: a
  controller \abs{FlowCtrl} and the tank itself \abs{Tank}.  The Tank
  has an in-port \abs{inDrain} and an out-port method
  \abs{outPort}. It has no controller, the \abs{run} method is empty.

  The controller's fields \abs{drain}, \abs{level} are its local
  copies of the state of the tank.  The controller method first
  updates \abs{level}, decides on the state of \abs{drain}, then
  pushes the (possibly changed) state of \abs{drain} to the tank.  No
  time passes in the controller after suspension which ensures that
  the copied fields are synchronized at the end of the round. As the
  \abs{Tank}'s fields are not directly accessible to the
  \abs{FlowCtrl} instance it is not possible to wait on the
  \abs{Tank}'s \abs{level}. Instead, every \abs{tick} seconds the
  controller is run.

  The \abs{Tank} interface specification declares an input requirement
  and a guarantee on returned values.  The precondition of the
  \abs{inDrain} method specification is a constraint on the input
  parameter.  The \abs{timed_requires} clause stipulates that the
  \abs{inDrain} method must be called at least once per second.
\end{example}

\subsection{Semantics of HABS}

HABS extends the SOS semantics for Timed ABS \cite{BjorkBJST13} and
only requires three small extensions:
\begin{inparaenum}[(i)]
\item include the physical behavior in the object state;
\item determine whether a differential guard holds and, if not, when
  it will at the earliest;
\item update the state whenever time passes.
\end{inparaenum}
These affect only expression evaluation and auxiliary functions. No
new SOS rule in addition to \cite{BjorkBJST13} is needed.

\paragraph{States.}

The state of an object has two parts:
\begin{inparaenum}[(i)]
\item a store $\sigma$, that maps (physical and non-physical) fields
  to values and the variables of the active process to values, and
\item $F$, the set of solutions of the ODE in its physical block.
\end{inparaenum}
A solution $f$ is a function from time to a store which only contains
the physical fields.  The state and the solutions are connected: for
each $f\in F$ and each physical field \abs{f} the following holds:
$f(0)(\xabs{f}) = \sigma(\xabs{f})$.

% For evaluation of expressions in a state $\sigma$ a function
% $\sem{\cdot}_\sigma$. As expressions are evaluated instantly, it is
% no necessary to refer to the solutions for evaluating them.

\paragraph{Semantics of Differential Guards.}

The semantics of an \abs{await g} statement is to suspend until the
guard holds, i.e.\ until $\sem{\xabs{g}}_\sigma^F$ evaluates to true.
A duration guard \abs{duration(e1,e2)} evaluates to true if
$\sem{\xabs{e1}}_\sigma^F \leq 0$.
% and future guard \abs{exp?} if the future stored in \abs{exp} is
% resolved\ektodo{Shall we even include futures?}.
%
Defining this extended timed semantics requires two operations on
guards: An extension of the \emph{evaluation function} that returns
true if the guard holds and the \emph{maximal time elapse}
$\mathit{mte}_\sigma^F$ returning the time $t$ that may maximally
elapse without the guard evaluating to true, or $\infty$ if this is
never the case.
We start with the semantics of expressions containing physical fields.

\begin{definition}[Semantics of Expressions with Physical
  Fields]\label{def:expr}
  Let $F$ be the set of solutions of an object $o$. Given a state
  $\sigma$ of $o$, we can check whether $F$ is a model of an
  expression \abs{e} at time $t$. Let $\xabs{f}_p$ be a physical field
  and $\xabs{f}_d$ a discrete field of $o$.  The semantics of fields
  $\xabs{f}_p$, $\xabs{f}_d$, unary operators \abs{!}, \abs{-} and
  binary operators
  $\sim\,\in\{$\abs{|},\,\abs{\&},\,\abs{>=},\,\abs{<=},\,\abs{+},\,\abs{-},\,\abs{*},\,\abs{/}$\}$
  is defined as follows:
  \begin{align*}
    \sem{\xabs{f}_p}_\sigma^{F,t} &= \cased{v & \text{if $\forall f\in F. v = f(t)(\xabs{f}_p)$} \\ \infty &\text{otherwise}}\qquad\sem{\xabs{f}_d}_\sigma^{F,t} = \sigma(\xabs{f}_d)  \\
    \sem{\xabs{-e}}_\sigma^{F,t} &= -\sem{\xabs{e}}_\sigma^{F,t}\qquad\sem{\xabs{!e}}_\sigma^{F,t} = \neg\sem{\xabs{e}}_\sigma^{F,t}\qquad\sem{\xabs{e} \sim \xabs{e'}}_\sigma^{F,t} = \sem{\xabs{e}}_\sigma^{F,t} \sim \sem{\xabs{e'}}_\sigma^{F,t}
  \end{align*}
\end{definition}

Outside differential guards, only the evaluation in the initial state
$\sem{\xabs{v}_c}_\sigma^{F,0}$ is needed and this expression is never
$\infty$.
% For integration, we identify the original evaluation
% $\sem{\cdot}_\sigma$ with $\sem{\cdot}_\sigma^{F,0}$.
Next we define $\mathit{mte}(\xabs{e})$: the \emph{maximal} time that
may elapse without missing an event is the \emph{minimal} time needed
by the system to evolve into a state where the guard holds. This
yields also the semantics of the guard itself.  We identify
$\sem{\cdot}^F_{\sigma}$ with $\sem{\cdot}^{F,0}_{\sigma}$.

\begin{definition}[Semantics of Differential Guards]\label{def:semant-diff-guards}
  Let $F$ be the set of solutions of object $o$ in state
  $\sigma$. Then we define:
  % we compute the needed time increase by the minimal time that is
  % needed until some element $F$ is a model for some expression
  % \abs{e}.
  \begin{align*}
    \mathit{mte}_\sigma^F(\xabs{diff e}) &= \argmin_{t\geq 0}~ \big(\sem{\xabs{e}}_\sigma^{F,t} = \text{\normalfont\text{true}}\big)
  \end{align*}
  A differential guard is evaluated to true if no time advance is
  needed:
  \begin{align*}
  \sem{\xabs{diff e}}_\sigma^{F,0} = \text{\normalfont\text{true}} &\iff \mathit{mte}_\sigma^F(\xabs{diff e}) = 0
  \end{align*}
\end{definition}
%The function $\mathit{mte}(\xabs{e})^F_\sigma$ is either $0$ or $\infty$.

If \abs{e} contains no continuous variable then the differential guard
semantics and the evaluation of expressions in Def.~\ref{def:expr}
coincides with condition synchronization and expression evaluation in
the standard ABS semantics \cite{JohnsenHSSS10}.

\paragraph{Time Advance.}

The characteristic feature of hybrid objects is that their (physical)
state changes when time advances, even when no process is active.
This is expressed in the semantics by a function
$\mathit{adv}(\sigma,t)$ which takes a state $\sigma$, a duration $t$,
and advances $\sigma$ by $t$ time units:
\[
  \mathit{adv}(\sigma,t)(\xabs{f}) =
  \cased{
    \sigma(\xabs{f}) & \text{ if \abs{f} is not physical} \\
    v & \text{ if }\forall f \in F.~v = f(t)(\xabs{f})}
\]
Hence, for non-hybrid Active Objects
$\mathit{adv}(\sigma,t) = \sigma$. Here the function is only needed to
modify the process pool of an object for scheduling, not its state,
and used exactly as in \cite{BjorkBJST13}.

%%% Local Variables:
%%% mode: latex
%%% TeX-master: "paper"
%%% End:

\section{Verification}\label{sec:translate}
\subsection{Overview}
\label{sec:overview}

With verification we mean essentially that an HAO satisfies its class
\emph{invariant} provided that the constraints expressed in the
preconditions are met. We make this precise now.
%
% In Ex.~\ref{ex:tank1} and Ex.~\ref{ex:tank2} the invariants were
% annotated to the class.  Additionally to the invariant, we also
% specify a precondition for the class parameters and conditions on the
% port methods.  The precondition for the class parameters restricts the
% possible starting values of the fields of an object, e.g., the
% precondition in Ex.~\ref{ex:tank1} expresses that the initial water
% level is within the set boundary.
%
A class specification is a tuple
$(\mathsf{inv}, \mathsf{pre}, \mathsf{TReq}, \mathsf{Req},
\mathsf{Ens})$, where \textsf{inv} is the class invariant, a \ddl
formula over the fields and parameters of the class and \textsf{pre}
is its precondition, a \ddl formula over the class parameters.
\textsf{TReq} is the set of timed preconditions for in-port methods:
\ddl formulas over a dedicated program variable with the method's name.
\textsf{Req} is the set of preconditions for in-port methods: \ddl
formulas over fields and the parameter of the methods.  \textsf{Ens}
is the set of postconditions for out-port methods: \ddl formulas over
a dedicated program variable with the method's name.

To verify a class, it is translated into a \ddl-formula that expresses
relative safety at any point in time and has the following form:
% is valid iff every object run that starts in a state satisfying the
% precondition of that object's class reaches only states that satisfy
% the class safety condition (at all points of time).  The proof
% obligation $\phi_\xabs{C}$ is
%
\begin{equation}
  \phi_\xabs{C}\;\equiv\;\texttt{precondition}_\xabs{C}\rightarrow\left[\left(\texttt{code}_\xabs{C};\;\texttt{plant}_\xabs{C}\right)^\ast\right]\texttt{safety}_\xabs{C}\label{eq:1}
\end{equation}

The \texttt{precondition}$_\xabs{C}$ is composed of \textsf{pre} and
restrictions on time and tick variables. As usual in controller
verification, the program consists of a control part
\texttt{code}$_\xabs{C}$, followed by the evolution of the continuous
behavior \texttt{plant}$_\xabs{C}$.  The \texttt{safety}$_\xabs{C}$
condition must hold after an arbitrary number of iterations. It
contains \textsf{inv}, preconditions of in-port methods of
referred objects, and postconditions of own out-port methods. The
following translation of a HABS class and its specification defines
formally how the placeholders are composed.

\subsection{Translation}
\label{sec:translation}

We need two operations on program sets $P$. The first constructs a
program that non-deterministically executes one of the arguments.  The
second constructs all programs that execute all arguments in some
order.
\begin{align*}
\sum P &= \sum \{p_1,\dots,p_n\} =p_1 + p_2 + \dots + p_n\\
\prod P &= \left\{ p_1;\ldots;p_{|P|} \mid \forall i,j \leq |P|.~ i \neq j \rightarrow {p}_i \neq {p}_j \wedge {p}_i, {p}_j \in P \right\}
\end{align*}

The translation has four phases:
\begin{inparaenum}[(i)]
\item provision of program variables,
\item code generation,
\item generation of the safety condition,
\item provision of ODEs and constraints.
\end{inparaenum}
We assume a fixed class \abs{C} with some technical constraints:
\begin{inparaenum}
\item\label{item:tech:1} If a controller writes to an in-port method
  of another object, then it reads from all out-port methods of the
  objects that occur in the precondition of that in-port method.
\item Every duration statement has two identical parameters.
\item In-port methods with a timed precondition are only called from
  timed controllers.
\item Local variable names are unique.
\end{inparaenum}
% Otherwise translation and verification fails.

\paragraph{Program Variables.}

For each field, parameter, and local variable in \abs{C} there is a
program variable with the same name.  For each method \abs{m} there is
a time variable $t_\xabs{m}\in\text{Time}$, for each in-port method
\abs{m} a tick variable $\mathit{tick}_{\xabs{m}}\in\text{Tick}$, both
type $\kyx{Real}$; $tick_{\xabs{m}}$ models the unknown time when an
in-port method is called next.

\begin{figure}[tb]
\noindent\scalebox{0.95}{\begin{minipage}{\textwidth}
\begin{align*}
  \transx{f} &\equiv f\text{ where $f$ is the \ddl program variable representing \xabs{f}}\\
  \transx{v} &\equiv v\text{ where $v$ is the \ddl program variable representing \xabs{v}}\\
  \trans(\xabs{e}_1~\mathit{op}~\xabs{e}_2) &\equiv \trans(\xabs{e}_1)~\mathit{op}~\trans(\xabs{e}_2)\\
  \trans(\xabs{if(e)}\{\mathtt{s}\}[\xabs{else}~\{\mathtt{s}\}]) &\equiv \kyx{if}~(\transx{e})~\kyx{then}~\trans(\mathtt{s})[\kyx{else}~\trans(\mathtt{s})]\\
  \trans(\xabs{while(e)\{s\}}) &\equiv \kyx{while}(\transx{e})\transx{s}\qquad\trans(\xabs{s}_1\xabs{;s}_2) = \trans(\xabs{s}_1)\xabs{;}\trans(\xabs{s}_2)\\
  \trans(\xabs{[T] v = e}) &\equiv \pumod{\transx{v}}{\transx{e}} \qquad \trans(\xabs{f = e}) \equiv \pumod{\transx{f}}{\transx{e}}\\
  \trans(\xabs{e!m()}) &\equiv \,\ptest{\mathit{true}}\qquad 
                         \trans(\xabs{f = e.m()}) \equiv \prandom{\transx{f}}\xabs{;}\ptest{\phi_{\xabs{m}}}\\
             &\hspace{-25mm}\text{ where $\phi_{\xabs{m}}$ is the postcondition of \abs{m}, with the method name replaced by $\trans(\xabs{f})$}%\\[-10mm]
\end{align*}
\end{minipage}}
\caption{Translation of statements and expressions}
\label{fig:trans1}
\end{figure}

\paragraph{Controller.}

The translation of ABS statements to hybrid programs is in
Fig.~\ref{fig:trans1}. We discuss the non-obvious rules:
Calls to in-port methods of other objects are mapped to
\ptest{\mathit{true}} (i.e.\ skip), because there is no effect on the
caller object.  A read from an out-port method is mapped to a
non-deterministic assignment, such that the read value adheres to the
postcondition of the called out-port method.  The translation of an
in-port or a controller has the form
\begin{equation}
  \kyx{if}~(\mathsf{check})~\kyx{then}~\{\mathsf{exec};\mathsf{cleanup}\}\label{eq:6}
\end{equation}

%\smtodo{briefly introduce translation notation; I would prefer a translation operator that is visually clearly distinct from the \dL operator $=$}
\begin{itemize}
\item For a timed controller \abs{m} with body \abs{await
    duration(e,e); s; this.m()}, $\mathsf{check}$ ensures that the
  correct duration passed and $\mathsf{cleanup}$ resets the clock:
  % We assume that the two parameters of \abs{duration} are equal for
  % verification, translation otherwise fails and the program is
  % rejected.
  \[\mathsf{check} \equiv t_{\xabs{m}} \doteq \transx{e} \qquad
    \mathsf{exec} \equiv \transx{s} \qquad \mathsf{cleanup} \equiv
    \pumod{t_{\xabs{m}}}{0}\]
\item For a differential controller \abs{m} with method body
  \abs{await diff e; s; this.m()}, $\mathsf{check}$ ensures the
  condition holds. No cleanup is needed:
  \[\mathsf{check} \equiv \transx{e} \qquad \mathsf{exec} \equiv \transx{s} \qquad \mathsf{cleanup} \equiv \ptest{\mathit{true}}\]
\item For an in-port method \abs{m} with method body \abs{this.f = v},
  precondition $\phi$ and timed precondition $\psi$, $\mathsf{check}$
  ensures the correct duration passed, $\mathsf{exec}$ chooses
  non-deterministically a value consistent with $\phi$, and
  $\mathsf{cleanup}$ does the same for a new duration consistent with
  $\psi$ (method name replaced by $\mathit{tick}_{\xabs{m}}$):
  \begin{align*}
    \mathsf{check} &\equiv t_{\xabs{m}} \doteq \mathit{tick}_{\xabs{m}} \qquad \mathsf{exec} \equiv \prandom{f};\ptest{\phi} \\
    \mathsf{cleanup} &\equiv \prandom{\mathit{tick}_{\xabs{m}}};\ptest{\mathit{tick}_{\xabs{m}} > 0};\ptest{\psi};\pumod{\mathit{t}_{\xabs{m}}}{0}
  \end{align*}
\end{itemize}
Let $M$ be the set of all translations of in-port methods and
controllers, then:
\begin{equation}
  \mathtt{code}_\xabs{C} \equiv \sum\prod M;\left(\sum M\right)^\ast\label{eq:3} 
\end{equation}

The controller \texttt{code}$_\xabs{C}$ first executes all controllers
in a non-deterministically chosen order, then allows each
controller/in-port to be run again.  The latter replicates eager ABS
behavior on satisfied guards: when a differential controller is
triggered and its guard still holds after its execution, then in ABS
the controller is run again.
We do not translate out-port methods and the \abs{run}
method. Out-port methods have no effect on an object state and their
post-condition is guaranteed at any point in time.  The \abs{run}
method only sets the system up and guarantees that every controller
has the chance to run before the plant.

\paragraph{Precondition and Safety Condition.}

The \texttt{precondition}$_\xabs{C}$ is \abs{C}'s precondition
\textsf{pre} plus restrictions on the time and tick variables: in the
beginning each clock starts at zero and the tick variables have an
unknown positive value. Additionally, all initalizations of physical
fields are added as equations. For example, \abs{Real r = param + 2}
results in $\xabs{r} \doteq \xabs{param}+2$.  The set of
all such initializations is \textsf{init}.
\begin{equation}
  \mathtt{precondition}_\xabs{C}\equiv
  \mathsf{pre}\wedge\bigwedge_{t\in\text{Time}}t\doteq0\wedge
  \bigwedge_{\mathit{tick}\in\text{Tick}}0<\mathit{tick}\wedge\bigwedge_{\phi\in\mathsf{init}}\phi\label{eq:2}
\end{equation}

The safety condition is \abs{C}'s invariant \textsf{inv} plus requires
clauses of the used in-port methods of other objects.
Recall technical constraint \ref{item:tech:1} above.  It ensures that
at the moment an in-port is called, the caller object has a correct
copy of the callee state.
% Concerning the timed preconditions, translation fails and the
% program is rejected if a method with such a precondition is called
% from a differential controller.
For a timed controller with guard \abs{duration(e,e)}, for each called
in-port method its timed precondition $\phi$ is added, but the method
name in $\phi$ replaced with $\trans(\xabs{e})$.  The set of 
modified preconditions is \textsf{TReq$'$}.  \textsf{Req$'$} are
preconditions of used in-port methods of other classes than \abs{C},
where the parameter is replaced by the field passed to it.
\textsf{Ens$'$} are the postconditions of all out-port methods of
\abs{C}, then:
\begin{equation}
  \mathtt{safety}_\xabs{C} \equiv \mathsf{inv} \wedge \bigwedge_{\psi\in \mathsf{Req'}}\psi\wedge
  \bigwedge_{\rho\in \mathsf{TReq'}}\rho\wedge\bigwedge_{\theta\in \mathsf{Ens'}}\theta\label{eq:5}
\end{equation}
% \paragraph{Assembly}
%The complete proof obligation has the following form.
%Let $M$ be the set of all translations of in-port methods and controllers.
%\begin{align*}
%&\texttt{precondition}\rightarrow\left[\left(\sum\prod M;\sum M;\texttt{plant}\right)^\ast\right]\texttt{safety}
%\end{align*}

\paragraph{Plant.}

The plant of a class \xabs{C} has the form
\begin{equation}
  \mathtt{plant}_\xabs{C} \equiv \sum\{(\mathtt{ode},\,\mathtt{ode}_t\,\&\,c) \mid c \in \mathcal{C}\}\enspace,
  % (\mathtt{ode},\,\mathtt{ode}_t\,\&\,c_1) + \dots+ (\mathtt{ode},\,\mathtt{ode}_t\,\&\,c_n)\enspace,
  \label{eq:4}
\end{equation}
where \texttt{ode} is the ODE from its physical block,
$\mathtt{ode}_t$ describes the clock variables, and the constraints
$c\in\mathcal{C}$ partition the domain of the physical fields.  The
boundaries of the subdomains overlap exactly where the differential
guards hold.\footnote{Expressions contain only \xabs{>=}, \xabs{<=},
  so weak complement ensures a boundary overlap.}  This models guards
as events in \ddl, following the modeling pattern described in
Sect.~\ref{sec:prelim}.
% TO%DO no idea
% Intuitively, the some ode in the plant is chosen and then followed
% for some (non-deterministically chosen) time but no longer than
% allowed by the guard. If the boundary is not reached yet, the
% controller is skipped.  If the boundary is reached, the
% corresponding differential controller is executed. By splitting the
% domain, it is not possible to jump over the condition of the
% differential guard, as this would require to switch into another ode
% of the plant.
To ensure that no differential guard is omitted, it is necessary that
no two differential guards share a program variable. This is not a
restriction, because two controllers can be merged with a disjunction
as in Expl.~\ref{ex:tank1}.

To define $\mathcal{C}$ let $e_1,\dots,e_m$ be the translations of
differential guards in the class and $\widetilde{e_i}$ the weak
complement of $e_i$.  Let $t_1,\dots,t_l$ be all time variables
introduced for timed controllers with $e_{t_i}$ the expression
in the \abs{duration} statement.  Let $pt_1,\dots,pt_k$ be all time
variables introduced for in-port methods and $\mathit{tick}_{pt_i}$
the associated tick variable.  We set
$\mathtt{ode}_t \equiv \{t_1' = 1,\dots,t_l' = 1,pt_1' = 1,\dots,pt_k' =
1\}$ and define:
\begin{align*}
  \mathcal{C} \equiv&\phantom{\cup} \left(\{e_1,\widetilde{e_1}\}\times\{e_2,\widetilde{e_2}\}\times \dots \times \{e_m,\widetilde{e_m}\}\right) \\
  &\cup \{t_1 \leq e_{t_i}\}_{i \leq l} \cup \{t_1 \geq e_{t_i}\}_{i \leq l} \cup \{pt_i \leq \mathit{tick}_{pt_i}\}_{i \leq n} \cup \{pt_i \geq \mathit{tick}_{pt_i}\}_{i \leq n}
\end{align*}

\paragraph{Remark.}

Instead of using the parameter precondition one can verify one
specific object created in a main block with a precondition that
characterizes it precisely.
---
Our translation generates a few basic invariants for handling hybrid
programs of the form $\alpha^*$. These include the safety condition,
the fact that the inner loop in \texttt{code} does not advance time,
and information on fields with final values.  These invariants are
usually not sufficient, but a starting point for manual specification.

\begin{theorem}\label{thm:1}
  Let $\mathsf{P}$ be a set of classes.  If for each
  $\xabs{C}\in\mathsf{P}$ the formula $\phi_\xabs{C}$ \eqref{eq:1} is
  valid, then for every main block that creates objects satisfying
  \textsf{pre} of $\xabs{C}$, in every reachable state all objects
  satisfy \textsf{inv} of $\xabs{C}$.
\end{theorem}

The main observations behind this theorem are:
\begin{inparaenum}
\item The \ddl program omits no events, because each event is at a
  boundary of two evolution constraints on a variable and no two
  events share a variable (each controller has its own time variable);
\item the evolution constraints cover all possible states, so no run
  is rejected, because the domain is too small;
\item each test in $\phi_\xabs{C}$ that discards runs, does so using a
  condition that has been proven.  For example, the test that discards
  all runs of an in-port method for inputs not satisfying its
  precondition is safe, because on the caller-side this condition is
  part of the safety condition.
\end{inparaenum}
The theorem also relies on technical constraint~\ref{item:tech:1} above and the fact that the recursive call is at the
end of a controller which guarantee that \emph{at this moment} the
caller copy of the callee's state is consistent with the callee's
actual state.

\subsection{Case Study}

\begin{figure}[tbh]
  %\centering
\includegraphics[scale=0.18]{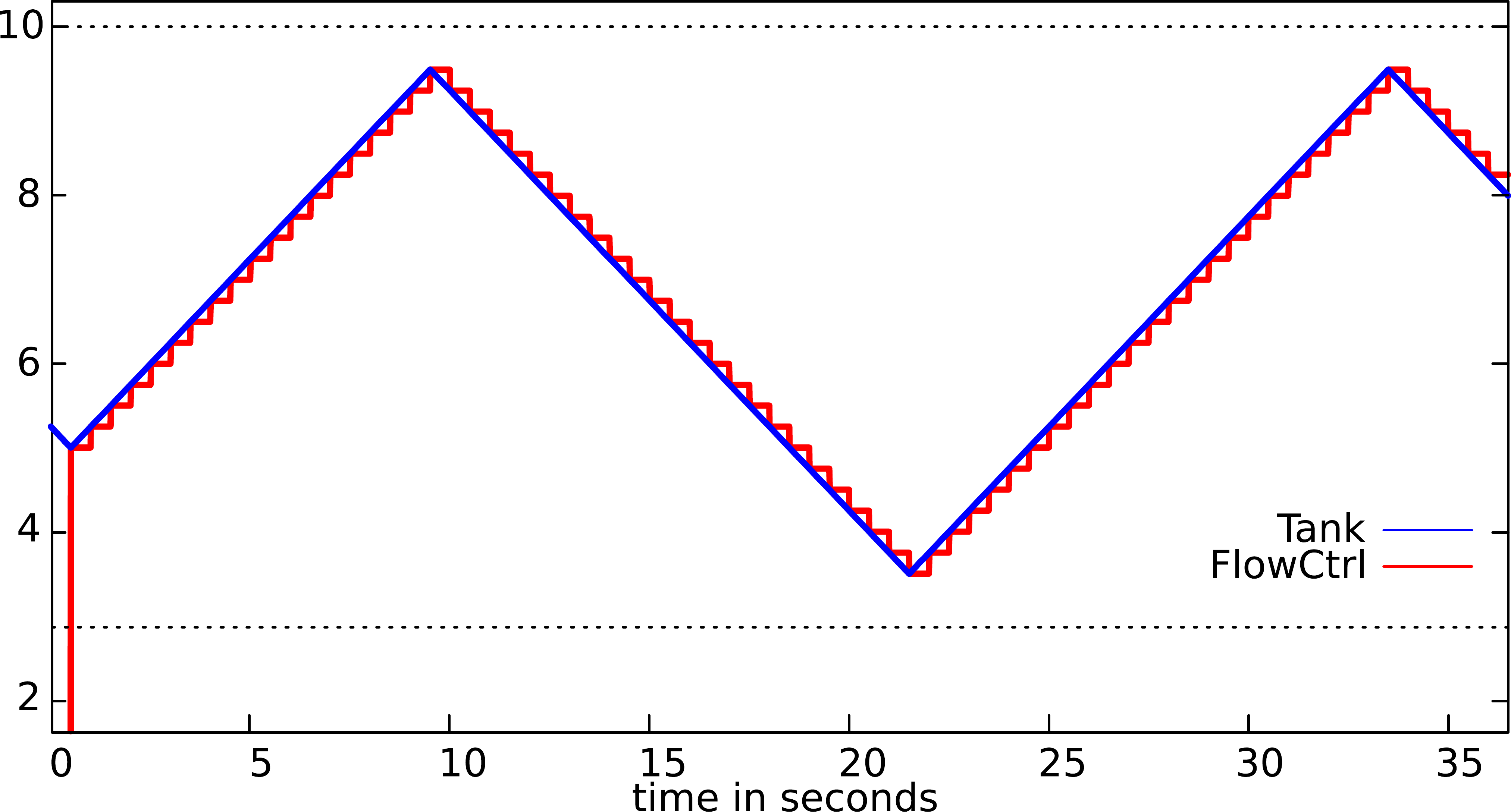}
\raisebox{0.1\height}{\includegraphics[scale=1.35]{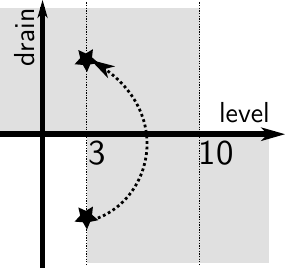}}
  \caption{Simulation of the two-object tank (l.) and events in the single-object tank (r.).}
  \label{fig:event}
\end{figure}

We illustrate the HABS to \KeYmaeraX\ translation (introduced in
Sect.~\ref{sec:translation} above) with the system in Fig.~\ref{fig:tank2}.
All files and mechanical proofs, the
translation and the simulator are available online under
\url{formbar.raillab.de/hybrid\_abs/}.
We first give the translation of the two-object water tank, whose behavior for an initial level of 5 is shown on the left of Fig.~\ref{fig:event}.

\paragraph{Tank.}

The in-port method of the \abs{Tank} class gives rise to a time
variable $\mathit{t}_{\xabs{inDrain}}$ and a tick variable
$\mathit{tick}_{\xabs{inDrain}}$.  Following \eqref{eq:2},
$\mathsf{precondition}_{\xabs{Tank}}$ is
\begin{align*}
  4 < \xabs{inVal} < 9 \wedge \mathit{t}_{\xabs{inDrain}} \doteq 0 &\wedge 0 < \mathit{tick}_{\xabs{inDrain}} \wedge \xabs{level} \doteq \xabs{inVal} \wedge \xabs{drain} \doteq -\frac{1}{2}
\end{align*}
The safety condition says the tank level stays in its limits and that
\abs{outLevel} adheres to its contract which happen to be identical. No
in-port methods of other classes are used, hence:
\[
  \mathsf{safety}_{\xabs{Tank}} \equiv 3 \leq \xabs{level} \leq 10
\]

The \abs{Tank} class has no controller method, so only the
\xabs{inDrain} method needs to be translated and we set
$\mathsf{code}_{\xabs{Tank}} \equiv \mathsf{p}(\mathsf{p})^\ast$ where
$\mathsf{p}\equiv\transx{inDrain}$, see~\eqref{eq:6}:

\noindent\scalebox{0.925}{\begin{minipage}{\textwidth}
\begin{align*}
  \mathsf{p} \equiv \kyx{if}~&( \mathit{t}_{\xabs{inDrain}} \doteq \mathit{tick}_{\xabs{inDrain}})~\kyx{then}\\
                       &\prandom{\xabs{drain}};\\
                       &\ptest{{\scriptsize -\frac{1}{2}} \leq \xabs{drain} \leq {\scriptsize\frac{1}{2}} \wedge (\xabs{drain} < 0 \rightarrow \xabs{level} \geq 3.5) \wedge (\xabs{drain} > 0 \rightarrow \xabs{level} \leq 9.5)};\\ 
                       &\prandom{\mathit{tick}_{\xabs{inDrain}}};\ptest{\mathit{tick}_{\xabs{inDrain}} > 0};\ptest{\mathit{tick}_{\xabs{inDrain}} < 1};\pumod{t_{\xabs{inDrain}}}{0}
\end{align*}
\end{minipage}}

\noindent Our implementation attempts to generate useful invariants. For the
example it proposes
$\mathsf{safety}_{\xabs{Tank}}\wedge\mathsf{old}(\xabs{inVal}) \doteq
\xabs{inVal}$ as a loop invariant for $(\mathsf{p})^\ast$. The second
conjunct is justified, because field \xabs{inVal} is not reassigned.
The $\mathsf{old}$ function yields the value of its argument before
the loop iteration. We do not attempt to validate the invariant, it
serves barely as a starting point for the interactive proof.
Obviously, $\mathsf{code}_{\xabs{Tank}}$ can be simplified to
$\mathsf{p}$, so an invariant is not necessary.

The plant~\eqref{eq:4} is based on the physical block and the new
clock variable (there are no differential guards), with the evolution
constraint split along the new time variable $t_\xabs{inDrain}$.  ODEs
of the form $v' = 0$ are default and omitted.
\begin{align*}
  \mathsf{plant}_{\xabs{Tank}} \equiv 
&\left(\{\xabs{level}' = \xabs{drain}, t_\xabs{inDrain}' = 1\,\&\,\mathit{t}_{\xabs{inDrain}} \leq \mathit{tick}_{\xabs{inDrain}} \}\right) + \\
&\left(\{\xabs{level}' = \xabs{drain}, t_\xabs{inDrain}' = 1\,\&\,\mathit{t}_{\xabs{inDrain}} \geq \mathit{tick}_{\xabs{inDrain}} \}\right)
\end{align*}

\paragraph{Timed Controller.}

The precondition~\eqref{eq:2} and plant~\eqref{eq:4} of \abs{FlowCtrl}
are straightforward. The latter is defined, even though the physical
block is not present.
\begin{align*}
  \mathsf{precondition}_{\xabs{FlowCtrl}} &\equiv 0 < \xabs{tick} < 1\\
\mathsf{plant}_{\xabs{FlowCtrl}} &\equiv \left(\{t_\xabs{ctrlFlow}'= 1\,\&\,t_{\xabs{ctrlFlow}} \geq \xabs{tick} \}\right) + \\
&\hspace{4mm}\left(\{t_\xabs{ctrlFlow}'\equiv 1\,\&\,(t_{\xabs{ctrlFlow}} \leq \xabs{tick} \}\right)
\end{align*}

The safety condition \eqref{eq:5} is the timed precondition of the
called \abs{inDrain} method and the class invariant (subsumed by the
precondition of \abs{inDrain}):

\noindent\scalebox{0.925}{\begin{minipage}{\textwidth}
\begin{align*}
&\mathsf{safety}_{\xabs{FlowCtrl}} \equiv \\
&{\scriptsize -\frac{1}{2}} \leq \xabs{drain} \leq {\scriptsize\frac{1}{2}} \wedge \xabs{tick} < 1 \wedge (\xabs{drain} < 0 \rightarrow \xabs{level} \geq 3.5) \wedge (\xabs{drain} > 0 \rightarrow \xabs{level} \leq 9.5)
\end{align*}
\end{minipage}}

\noindent Finally, the code $\mathsf{code}_{\xabs{FlowCtrl}}$ is defined as
$\mathsf{q}(\mathsf{q})^\ast$ with
\begin{align*}
\mathsf{q} \equiv \kyx{if}~&(t_\xabs{ctrlFlow} \doteq  \xabs{tick})~\kyx{then}~\\
&\prandom{\xabs{level}};~\ptest{3\leq\xabs{level}\leq10};~\kyx{if}~(\xabs{level}\leq 3.5)~\kyx{then}~\{\pumod{\xabs{drain}}{\frac{1}{2}}\}; \\
&\kyx{if}~(\xabs{level}\geq 9.5)~\kyx{then}~\{\pumod{\xabs{drain}}{-\frac{1}{2}}\};~\pumod{t_\xabs{ctrlFlow}}{0}
\end{align*}

\paragraph{Differential Controller.}

We give the translation of the tank \xabs{CSingleTank} from
Fig.~\ref{fig:tank1} to illustrate the handling of differential
controllers.  Its precondition and safety condition are again
straightforward:
\begin{align*}
  \mathsf{precondition}_{\xabs{CSingleTank}} &\equiv 4 < \xabs{inVal} < 9\\ 
  \mathsf{safety}_{\xabs{CSingleTank}} &\equiv 3 \leq \xabs{level} \leq 10 
\end{align*}

The plant and code interact. The plant separates the evolution domain
into two parts, with the guard of the differential controller (the
white areas in Fig.~\ref{fig:event}) defining their boundary. The gray
areas are larger than the safe region defined by \abs{3 <= level <=
  10}. This is necessary: If we used simply the complement of the safe
region \abs{level <= 3 | level >= 10} as a guard and are in a program
state at the boundary (the lower of the states indicated with a star
on the right in Fig.~\ref{fig:event}), then the controller changes the state as
shown by the arrow. But if the next state is \emph{on the boundary},
then the guard is triggered, the controller runs again, and so on,
%without physical time having a chance to advance. The guard chosen in
without physical time being able to advance. The guard in
Fig.~\ref{fig:tank1} ensures that after the controller has run, the
state is not on the boundary anymore.  This behavior is visualized by
our implementation, see the right part of Fig.~\ref{fig:tank1}.  The
code has the form
$\mathsf{code}_{\xabs{CSingleTank}} \equiv \mathsf{r}(\mathsf{r})^{\ast}$,
with \textsf{r} defined as follows:
\begin{align*}
  \mathsf{r} &\equiv \kyx{if}~(\xabs{level}\leq3\wedge\xabs{drain}\leq0)\vee(\xabs{level}\geq10\wedge\xabs{drain}\geq0)~\kyx{then}\\
&\hspace{20mm}\kyx{if}~(\xabs{level}\leq3)~\kyx{then}~\pumod{\xabs{drain}}{\frac{1}{2}}~\kyx{else}~\pumod{\xabs{drain}}{-\frac{1}{2}}\\
&\mathsf{plant}_{\xabs{CSingleTank}} \equiv\\
&\left(\{ \xabs{level}' = \xabs{drain}\,\&\,(\xabs{level}\leq 3 \wedge \xabs{drain} \leq 0) \vee (\xabs{level} \geq 10\wedge \xabs{drain} \geq 0) \}\right) + \\
&\left(\{\xabs{level}' = \xabs{drain}\,\&\,(\xabs{level}\geq 3 \wedge \xabs{drain} \leq 0) \vee (\xabs{level} \leq 10\wedge \xabs{drain} \geq 0) \}\right) 
\end{align*}

%%% Local Variables:
%%% mode: latex
%%% TeX-master: "paper"
%%% End:

%%% Local Variables:
%%% mode: latex
%%% TeX-master: "paper"
%%% End:

\section{Simulation}\label{sec:impl}
%\begin{floatingfigure}[l]{0.5\textwidth}
%\end{floatingfigure}

The implementation of HABS is based on the ABS compiler
\cite{WongAMPSS12}, can also be used stand-alone, and is not subject
to the restrictions imposed for translation to \ddl\ given in
Fig.~\ref{fig:syntax}. It allows unrestricted use of differential
guards and includes ABS features such as synchronization via futures,
a module system, product lines, and abstract data types.

% In this section we explain how HABS Model are simulated. As
% mentioned previously ABS includes a various number of tools. One of
% them is the Erlang backend which simulates ABS models by translating
% them into Erlang and provides a runtime library that implements key
% features of ABS. However, Erlang is not able to handle continuous
% behaviour, because it cannot solve OEDs. In order to describe
% continuous behaviour we use the computer algebra system
% Maxima\cite{Haager2014}. It can be called from the command line and
% runs on many operating systems. Further, Maxima is used to visualize
% the simulation as a graph.

% The continuous behaviour of a HAO is described in the physical block
% as a set of physical fields with ODEs above them. The discrete
% behaviour is described by the condition of the \abs{await} statement
% and specifies when the process can be rescheduled. Internally, the
% time of the next discrete event is calculated and the time will be
% increased step by step.

The ODEs of a class cannot be changed at runtime and are represented
as a string in the class table.  To compile differential guards
correctly one needs to compute $\mathit{mte}^F_\sigma(\xabs{diff e})$
(Def.\ \ref{def:semant-diff-guards}). To compute $F$, the ODEs and the
current state of the physical fields are passed to
Maxima~\cite{maxima} as an \emph{initial value problem}.  The solution
is an equation system or an error. Currently, the simulator does not
support non-unique solutions or non-solvable ODEs.

Afterwards, Maxima is invoked again, this time with a
\emph{minimalization problem}: it minimizes the time $t$ with the
equation system representing $F$ as the constraints.  The result is
then handled in the same way as a parameter to a timed guard by the
runtime system.  Once time has passed and the suspended process is
reactivated, the physical fields are updated according to $F$.
To implement the time advance function $\mathit{adv}$, if the state of
the object changes any physical field, the above procedure is executed
again for every currently suspended differential guard to accommodate
the result.

The output files used to visualize a program execution are of the form
$t_1, F_{1}$, $t_{1}, F_{2}$, $t_{2}, \dots, F_{n}, t_{n}$.  Here
$t_i$ are the points in time where the object schedules a process and
$F_i$ the function describing its physical behavior in the previous
suspended state.  Each time a differential guard is reactivated, not
only its state is updated, but the solution $F_{i+1}$ and the
reactivation time $t_{i+1}$ are written to the output.  Each object
has its own output file.
% For example the evolution of the physical fields is described from
% time $t_{0}$ until time $t_{1}$ with the function $f_{1}(t)$ and
% then the function $f_{2}(t)$ and following functions will be used in
% an analogous way.
A Python script translates output files into a discrete dynamic graph
in Maxima format which in turn calls \textsf{gnuplot} that is
responsible for creating the graph.  The output of one object is shown
in Fig.~\ref{fig:tank1} and an overlay of two objects in
Fig.~\ref{fig:event}. The latter shows how the state of the discrete
controller lags behind the physical state of the tank.

\section{Related \& Future Work, Conclusion}\label{sec:conclusion}
\paragraph{Related Work.}

% Recent efforts
% \cite{DBLP:journals/sttt/MullerMRSP18,DBLP:conf/acsd/LunelBT17}
% introduce component-based modeling and verification techniques for
% hybrid systems in \dL, with a focus on splitting verification
% efforts into more manageable pieces.
Recent efforts
\cite{DBLP:journals/sttt/MullerMRSP18,DBLP:conf/acsd/LunelBT17}
introduce component-based modeling and verification techniques for
hybrid systems in \dL to split the verification task into manageable
pieces.  Integrated tools such as Ptolemy \cite{ptolemy14} emphasize
timing aspects, signals, and data flow between heterogeneous models
and their simulation.  Our work complements these efforts with
user-friendly modeling constructs in a uniform modeling language,
validation by simulation, and modular formal verification along system
aspects
% (ABS for distributed system analysis, \dL for hybrid systems
% verification)
through translation from HABS to \dL.

Translation between languages for hybrid systems so far is centered on
hybrid automata as a unifying notation
\cite{DBLP:conf/hybrid/BeekRSR07,DBLP:conf/hybrid/BakBJ15}.  Others
focus on the discrete fragment \cite{DBLP:conf/iccps/GarciaMP19}.  Our
translation from HABS to \dL translates full hybrid systems models
written in a programming language, including annotations (e.g.,
starting and invariant conditions); it is soundly based on the formal
semantics of HABS and \dL.

Hybrid systems validation through simulation is approached with
translation to Stateflow/Simulink
\cite{DBLP:journals/sttt/BakBBJNS19}; with a combination of
discrete-event and numerical methods
\cite{DBLP:conf/hybrid/BrooksLLNW15}; with threat models for security
analyses \cite{DBLP:journals/pieee/KoutsoukosKLNPV18}; and with
co-simulation between control software and dedicated physics
simulators
\cite{DBLP:journals/simpra/ZhangEKPKS14,DBLP:conf/models/CremonaLBLMT18}.
Here, we focus on safety verification, the distributed aspect of HABS
models, and take a pragmatic first step for simulating continuous models. % (with polynomial solutions).

Hybrid Rebeca~\cite{hybreb} proposes to embed hybrid automata directly
into the actor language Rebeca.
% To do so, hybrid automata are encoded directly into special physical
% classes that run separatly from software actors.  Actors are more
% restricted than Active Objects and Hybrid Rebeca supports no
% \abs{await} statement outside of its hybrid automata and relies on
% call-backs to realize events.
In contrast to HABS, no simulation is available and verification is
not object-modular, because the whole model is translated into a
single hybrid automaton. Because of this, a number of boundedness
constraints have to be imposed.  The verification backend of Hybrid
Rebeca does not support non-linear ODEs (our examples are linear, but
HABS, KeYmaera X, and Maxima, support non-linear ODEs; HABS models
with non-linear ODEs are found in the online supplement).  Hybrid
automata can be encoded in HABS, see Fig.~\ref{fig:tank1}.

Only one model~\cite{KHS18} uses Active Objects for a cyber-physical
system, with the larger part of the model simulating the differential
guards introduced here.

\paragraph{Future Work.}

We plan to extend the verification backend to more liberal
differential guards and fields. We want to integrate the verification
of hybrid properties described here with the verification of
computation-heavy Active Objects, as implemented in the KeY-ABS tool
\cite{DinBH15}.
% We furthermore plan to investigate how the static analyses for ABS
% can be extended for HABS.
We also plan to implement approximating the simulation of non-solvable
ODEs and to perform a large case study.

\paragraph{Conclusion.}

Concurrent hybrid systems are not only difficult to \emph{verify}
formally, it is equally hard to \emph{validate} a formal model of
them.  Both activities have conflicting demands, so we propose a
translation-based approach: modeling is guided by patterns over hybrid
programs and class specifications in HABS, a hybrid extension of the
concurrent AO language ABS. These are automatically decomposed and
translated in a semantics-preserving manner (Thm.~\ref{thm:1}) into
sequential proof obligations of the verification-oriented differential
dynamic logic \ddl and discharged by the hybrid theorem prover
\KeYmaeraX.
We illustrated the viability of our approach by two case studies that
feature many complications: concurrent behavior, possible
non-termination, correctness depending on timing constants,
multi-dimensional domain, time lag in sensing, etc.
Asynchronous method calls and differential guards keep the hybrid AO
models succinct and intuitive, while simulation and visualization
support their validation.

\iffalse
%
We presented Hybrid Active Objects and the Hybrid Abstract Behavioral
Specification language that implements them and demonstrated with a
water tank case study how they can be used to model, specify, verify
and simulate distributed cyber-physical systems.  Due to their
conceptual closeness to programs, HAOs are more natural than classical
low-level modeling approaches. Yet, these low-level approaches are
integrated into the tool-suite of HABS by an appropriate translation.
By extending the ABS language, HABS automatically gets support for
features such as traits, programming-to-interface, product lines and a
module system, that further increase usability.
%
\fi

%%% Local Variables:
%%% mode: latex
%%% TeX-master: "paper"
%%% End:

\clearpage

\bibliographystyle{splncs04}
\bibliography{ref}
\clearpage
\appendix
For convenience, we give the semantics of Timed ABS here.
The content of this section is a reproduction of Section 4 in~\cite{BjorkBJST13}, with kind permission of the authors.

\section{Semantics}
Fig.~\ref{fig:abs-comps:bnf} gives the object layer of Timed ABS, which is extended in Fig.~\ref{figure:config} to the runtime syntax.
\begin{figure*}[t]
\myttsize
$$\begin{array}{cl}
      \begin{array}[t]{l}
        \emph{Syntactic categories.}\\
        C, I, m \textrm{ in } \TYPE{Name}\\       
       g \textrm{ in } \TYPE{Guard}\\
        s \textrm{ in } \TYPE{Stmt}\\
a  \textrm{ in } \TYPE{Annotation}
     \end{array}
      \;\;
      &
      \begin{array}[t]{rrl}
        \multicolumn{3}{l}{\emph{Definitions.}}\\
        \mathit{IF}&::=&
        \key{interface}\ I\,\{\,\option{\many{Sg}}\,\}\\
        CL &::=& \option{\texttt{[}a\texttt{]}}\;\key{class}\,C\,
        \option{(\many{T\ x})}\, \option{\key{implements}\,\many{I}}\,\{\,\option{\many{T\ x};}\, \many{M}\}\\
       \textsl{Sg} &::=& T\ m\ (\option{\many{T\ x}})\\
       \textsl{M} &::=& \option{\texttt{[}a\texttt{]}}\;\textsl{Sg}\ \{ \option{\many{T\ x};}\ s\, \}\,\\   
a &::=& \texttt{Deadline:}\; d \asep \texttt{Cost:}\; d \asep \texttt{Critical:}\; b \asep \texttt{Scheduler}: e \asep a,a \\
        g&::=& b \asep x? \asep \key{duration}(d,d) \asep g\land g \\
      s&::=& s;s \asep \key{skip} \asep \key{if}\ b\ \{\, s\, \}\, \option{\,\key{else}\, \{\, s\, \}}
        \asep \key{while}\ b\, \{\, s\, \}  \asep \key{return}\ e    \\
       &|& \option{\texttt{[}a\texttt{]}}\;x=rhs \asep
      \key{suspend} \asep \key{await}\ g \asep
      \key{duration}(d,d) \\
       rhs &::=& e \asep \key{new} \ C\,(\many{e}) \asep 
       e.\key{get} \asep o!m(\many{e}) 
     \end{array}\end{array}$$
\caption{\label{fig:abs-comps:bnf}Syntax for the concurrent object
  level.}
\end{figure*}

\begin{figure}[t]
\myttsize 
\renewcommand{\arraystretch}{0.9} 
$$\begin{array}{rcl@{\hspace{10pt}}rcl}
e &::=& \key{case2}\ v\ \{\many{br}\}\asep \ldots&
v & ::= & o \asep f\asep \ldots\\
s & ::= & \key{duration2}(d_1,d_2)\asep \ldots\\
\nt{cn} & ::= & \epsilon \asep \nt{obj} \asep \nt{msg} \asep \nt{fut} \asep  \nt{cn}~\nt{cn}&
\nt{tcn} & ::= & \nt{cn}~\text{clock}(t)\\
\nt{fut} & ::= & f \asep \text{fut}(f,v)&
\sigma & ::= & x\mapsto v \asep \sigma\circ\sigma\\
\nt{obj} & ::= & \text{ob}(o,e,\sigma,pr,q) &
pr&::=& \{\sigma | s\} \asep \text{idle}\\
\nt{msg} & ::= &m(o,\many{v},f,d,c,t)&
q & ::= & \epsilon \asep pr \asep q \circ q 
\end{array}$$
\caption{Runtime syntax; here, $o$ and $f$ are
  object and future identifiers, $d$ and $c$ are the deadline and cost
  annotations.}
\label{figure:config}
\end{figure}

\emph{Configurations} $\mathit{cn}$ are sets of objects, invocation
messages, and futures.  A \emph{timed configuration} $\nt{tcn}$ adds a
global clock $\nt{clock}(t)$ to a configuration (where $t$ is a value
of type \TYPE{Time}). The global clock is used to record arrival and
finishing times for processes. Timed configurations live inside curly
brackets; thus, in $\{cn\}$\!, $cn$ captures the \emph{entire} runtime
configuration of the system. The associative and commutative union
operator on (timed) configurations is denoted by whitespace and the
empty configuration by $\none$.

An \emph{object} $\nt{obj}$ is a term $ob(o,e,\sigma,\textit{pr},q)$
where $o$ is the object's identifier, $e$ is an expression of type
\TYPE{Process} representing a \emph{scheduling policy}, $\sigma$ a
substitution representing the object's fields, $pr$ is an (active)
process, and $q$ a \emph{pool of processes}.  A \emph{substitution}
$\sigma$ is a mapping from variable names $x$ to values $v$.  For
substitutions and process pools, concatenation is denoted by
$\sigma_1\circ \sigma_2$ and $q_1 \circ q_2$, respectively.

In an \emph{invocation message} $m(o,\many{v},f,d,c,t)$, $m$ is the
method name, $o$ the callee, $\many{v}$ the call's actual parameter
values, $f$ the future to which the call's result is returned, $d$ and
$c$ are the provided deadline and cost of the call, and $t$ is a time
stamp recording the time of the call.  A \emph{future} is either an
identifier $f$ or a term $\mathit{fut}(f,v)$ with an identifier $f$
and a reply value $v$.  For simplicity, classes are not represented
explicitly in the semantics, but may be seen as static tables of
object layout and method definitions.

\paragraph{Processes and Process Lifting.}
A \emph{process} $\{\sigma |s\}$ consists of a substitution $\sigma$
of local variable bindings and a list $s$ of statements, or it is
\textit{idle}. By default, the local variables of a process include
the variables \texttt{method} of type \TYPE{String}, \texttt{arrival}
of type \TYPE{Time}, \texttt{cost} of type \TYPE{Duration},
\texttt{deadline} of type \TYPE{Duration}, \texttt{start} of type
\TYPE{Time}, \texttt{finish} of type \TYPE{Time}, \texttt{critical} of
type \TYPE{Bool}, \texttt{value} of type \TYPE{Int}, and
\texttt{destiny} of type \TYPE{Name}. Consequently, we can define a
function \emph{lift} which transforms the runtime representation of a
process into the Timed ABS datatype of processes and a function
\emph{select} which returns the process corresponding to a given
process identifier in a process queue, as follows: 
{\myttsize
$$\begin{array}{l}
\begin{array}{ll}
%\textit{lift}(\texttt{idle})&=\textit{Idle}\\
\textit{lift}(\{\sigma| s\})&=\texttt{Proc}
 (\sigma(\texttt{destiny}),\sigma(\texttt{method}),
\sigma(\texttt{arrival}),\\
&\quad\;\sigma(\texttt{cost}),\sigma(\texttt{deadline}),
\sigma(\texttt{start}),
\sigma(\texttt{finish}),\\
&\quad\;\sigma(\texttt{crit}),
\sigma(\texttt{value}))
\end{array}\\
\textit{select}(\textit{pid}, \varepsilon)=\text{idle}\\
\textit{select}(\textit{pid}, \{\sigma|s\}\append q)=
\left\{\begin{array}{ll}
\{\sigma|s\}&\text{if}~\sigma(\texttt{destiny})=\textit{pid}\\
\textit{select}(\textit{pid}, q)~&\text{otherwise}
\end{array}\right.
\end{array}$$}%
The value of \texttt{destiny} is guaranteed to be unique, and is
% can be 
used to identify processes at
the Timed ABS level.

\begin{figure}[t]
\myttsize
$$
\begin{array}{l}
\eval{b}{\sigma}=b\\
\eval{x}{\sigma}=\sigma(x)\\
\eval{v}{\sigma}=v\\
\eval{Co(\many{e})}{\sigma}=Co(\eval{\many{e}}{\sigma})\\
\eval{\key{deadline}}{\sigma}=\sigma(\texttt{deadline})
\quad\\
\eval{\mathit{fn} (\many{e})}{\sigma}=
\left\{ 
\begin{array}{ll}
\eval{e_{\mathit{fn}}}{\many{x}\mapsto\many{v}}  & \text{if}\ \many{e}=\many{v}\\
\eval{\mathit{fn} (\eval{\many{e}}{\sigma})}{\sigma}  & \text{otherwise}
\end{array}\right.
\\\quad\\
\eval{\key{case}\ e\ \{\many{br}\}}{\sigma}=\eval{\key{case2}\ \eval{e}{\sigma}\ \{\many{br}\}}{\sigma}\\
\eval{\key{case2}\ t\ \{p\Rightarrow e; \many{br}\}}{\sigma}=
\left\{ 
\begin{array}{ll}
\eval{e}{\sigma\circ\mathit{match}(p,t)}\quad & \text{if}\ \mathit{match}(p,t)\neq\bot\\
\eval{\key{case2}\ t\ \{\many{br}\}}{\sigma} & \text{otherwise}
\end{array}\right.\\[-10pt]
\end{array}$$
\caption{The evaluation of functional expressions. \label{fig:eval-func}}
\end{figure}

\subsection{A Reduction System for Expressions}
\label{app:eval}

The strict evaluation $\eval{e}{\sigma} $ of functional expressions
$e$, given in Fig.~\ref{fig:eval-func}, is defined inductively over
the data types of the functional language and is mostly standard,
hence this subsection only contains brief remarks about some of the
expressions.  Let $\sigma$ be a substitution which binds the name
\texttt{deadline} to a duration value.  For every (user-defined)
function definition {\myttsize $$\key{def}\ T\ \mathit{fn}(\many{T \
    x})= e_{\mathit{fn}},$$} the evaluation of a function call
$\eval{\mathit{fn}(\many{e})}{\sigma}$ reduces to the evaluation of
the corresponding expression
$\eval{e_{\mathit{fn}}}{\many{x}\mapsto\many{v}}$ when the arguments
$\many{e}$ have already been reduced to ground terms $\many{v}$.
(Note the change in scope. Since functions are defined independently
of the context where they are used, we here assume that the expression
$e$ does not contain free variables and the substitution $\sigma$ does
not apply in the evaluation of $e$.)  In the case of pattern matching,
variables in the pattern $p$ may be bound to argument values in
$v$. Thus the substitution context for evaluating the right hand side
$e$ of the branch $p\Rightarrow e$ extends the current substitution
$\sigma$ with bindings that occurred during the pattern matching.  Let
the function $\mathit{match}(p,v)$ return a substitution $\sigma$ such
that $\sigma(p)=v$ (if there is no match, $\mathit{match}(p,v)=\bot$).

\subsection{A Transition System for Timed Configurations}
\emph{Evaluating Guards.}  Given a substitution $\sigma$ and a
configuration $cn$, we lift the evaluation function for functional
expressions and denote by $\geval{g}{\sigma}{cn}$ a evaluation
function which reduces guards $g$ to data values (the state
configuration is needed to evaluate future variables). Let $\geval{g_1
  \land g_2}{\sigma}{cn}= \geval{g_1}{\sigma}{cn} \land
\geval{g_2}{\sigma}{cn}$,
$\geval{\key{duration}(b,w)}{\sigma}{cn}=\eval{b}{\sigma}\leq 0$,
$\geval{x?}{\sigma}{cn}=\text{true}$ if $\eval{x}{\sigma}=f$ and
$\textit{fut}(f,v)\in cn$ for some value $v$ (otherwise $f\in \nt{cn}$
and we let $\geval{x?}{\sigma}{cn}=\text{false}$), and
$\geval{b}{\sigma}{cn}=\eval{b}{\sigma}$.

\emph{Auxiliary functions.}  If $T$ is the return type of a method $m$
in a class $C$, we let $\textit{bind}(m,o,\many{v},f,d,b,t)$ return a
process resulting from the activation of $m$ in the class of $o$ with
actual parameters $\many{v}$, callee $o$, associated future $f$,
deadline $d$, and criticality $b$ at time $t$.  If binding succeeds,
this process has a local variable \texttt{destiny} of type
$\TYPE{fut}\langle T\rangle$ bound to $f$, the method's formal
parameters are bound to $\many{v}$, and the reserved variables
\texttt{deadline} and \texttt{critical} are bound to $d$ and $b$,
respectively. Furthermore, \texttt{arrival} is bound to $t$ and
\texttt{cost} to $\eval{e}{\many{x}\mapsto\many{v}}$ (or to the
default $0$ if no annotation is provided for the method).  The
function $\textit{atts}(C,\many{v},o)$ returns the initial state of an
instance of class $C$, in which the formal parameters are bound to
$\many{v}$ and the reserved variables $\texttt{this}$ is bound to the
object identity $o$.  The function $\textit{init}(C)$ returns an
activation of the \emph{init} method of $C$, if defined. Otherwise it
returns the \emph{idle} process.  The predicate $\textit{fresh}(n)$
asserts that a name $n$ is globally unique (where $n$ may be an
identifier for an object or a future).

\begin{figure*}[t!]
\scalebox{0.725}{\begin{minipage}{\textwidth}
\myttsize
$$\begin{array}{c}
\nredrule{Skip}
{\text{ob}(o,p,a, \{l \mid \key{skip};s\}, q)\\\to \text{ob}( o,p,a, \{l \mid s\}, q)}
\qquad
\ntyperule{Assign1}{x\in\,\textit{dom}(l)}
{\text{ob}(o,p,a, \{l \mid x=e;s\}, q)\\
\to \text{ob}(o,p, a, \{l[x\mapsto\eval{e}{{a}\circ{l}}] \mid s \}, q)}
\qquad
\ntyperule{Assign2}{x\not\in\,\textit{dom}(l)}
{\text{ob}(o,p,a, \{l \mid x=e;s\}, q)\\
\to \text{ob}(o,p, a[x\mapsto\eval{e}{{a}\circ{l}}], \{l \mid s \}, q)}
\\\\
\nredrule{Suspend}{\text{ob}(o,p,a, \{l\mid \key{suspend};s\}, q)\\
\to \text{ob}(o,p, a, \text{idle}, \{{l}\mid s\}\append q)}
\qquad
\ntyperule{Await1}{\geval{e}{a\circ l}{\textit{cn}}\\[-9pt]}
{\{\text{ob}(o,p,a, \{l\mid \key{await}\ e;s\}, q)\ \textit{cn}\}\\
\to \{\text{ob}(o,p, a, \{l\mid s\}, q)\ \textit{cn}\}}
\qquad
\ntyperule{Await2}{\neg \geval{e}{a\circ l}{\textit{cn}}\\[-9pt]}
{\{\text{ob}(o,p,a, \{l\mid \key{await}\ e;s\}, q)\  \textit{cn}\}\\
\to \{\text{ob}(o,p,a,   \{l\mid \key{suspend};\key{await}\ e;s\}, q)\  \textit{cn}\}}
\\\\
\ntyperule{Schedule}
{q'=\textit{ready}(q,a,cn)\quad\;\;\; pr=\textit{select}(pid,q)\\
q'\neq\emptyset\quad pid=\eval{procid(p)}{a[\texttt{queue}\mapsto\textit{liftall}(q')]}\\[-9pt]}
{\{\text{ob}(o,p,a, \text{idle}, q)\  \textit{cn}\}\\
\to \{\text{ob}( o,p,a, pr, (q\setminus pr))\ \textit{cn}\}}
\qquad
\ntyperule{Activation}{q'=\textit{bind}(m,o,\bar{v},f,d,b,t)\append q}
{\text{ob}(o,p, a, pr, q)\\  m(o,\bar{v},f,d,b,t)\\
\to \text{ob}(o,p, a, pr,q')}
\qquad
\ntyperule{New-Object}{an = \texttt{Scheduler:}\: p'\quad \textit{fresh}(o')\\
pr=\textit{init}(C) \quad a'=\textit{atts}(C,\eval{\many{e}}{a\circ l},o', c)\\[-9pt]}
{\text{ob}(o,p,a,\{l| [an]\;x=\key{new}\ C(\many{e});s\},q) 
\\\to \text{ob}(o,p,a,\{l|x=o';s\},q)\\
  \text{ob}(o',p',a',pr,\emptyset)}
\\\\
\ntyperule{Tick}{0 < d \leq \textit{mte}(cn)}
{\{cn\}\to \{\textit{adv}(cn,d)\}}
\qquad
\ntyperule{Duration1}{d_1=\eval{e_1}{{a}\circ{l}} \quad d_2=\eval{e_2}{{a}\circ{l}}\\[-9pt]}
{\text{ob}(o,p, a, \{l \mid \key{duration}(e_1,e_2);s\}, q)\\
\to \text{ob}(o,p, a, \{l \mid \key{duration2}(d_1,d_2);s\}, q)}
\qquad
\ntyperule{Duration2}{d_1 \leq 0}
{\text{ob}(o,p, a, \{l \mid \key{duration2}(d_1,d_2);s\}, q)\\
\to \text{ob}(o,p, a, \{l \mid s\}, q)}
\\\\
\ntyperule{Async-Call}{\textit{fresh}(f)\quad an= \texttt{Deadline:}\:d, \texttt{Critical:}\:b}{
\text{ob}(o,p, a, \{l \mid [an]\;x:=e!m(\many{e});s\}, q)\ \text{clock}(t)\\
\to \text{ob}(o,p, a, \{l \mid x:=f);s\}, q)\ \text{clock}(t)\\
m(\eval{e}{a\circ l}, \eval{\many{e}}{a\circ l}, f,d,b,t)\ f}
\qquad
\ntyperule{Return}{f={l}(\texttt{destiny})}
{\text{ob}(o,p,a, \{l\mid \key{return}(e);s\}, q)\ \text{clock}(t)\ f\\
\to \text{ob}(o,p,a,\{l\mid \texttt{finish}=t\}, q)\ \text{clock}(t)\ \text{fut}(f, \eval{e}{{a}\circ {l}})}
\\\\
\ntyperule{Read-Fut}{f=\eval{e}{a\circ l}\\[-9pt]}
{\text{ob}(o,p,a,\{l\mid x=e.\key{get};s\}, q)\ \text{fut}(f,v)\\
\to \text{ob}(o,p,a,\{l\mid x=v;s\}, q)\ \text{fut}(f,v)}
\qquad
\ntyperule{Cond1}{\eval{e}{a\circ l}}
{\text{ob}(o,p,a,\{l|\key{if}\ e\ \{ s_1\}\ \key{else}\ \{ s_2\};s\},q)\\
\to \text{ob}(o,p,a,\{l|s_1;s\},q)}
\qquad
\ntyperule{Cond2}{\neg \eval{e}{a\circ l}}
{\text{ob}(o,p,a,\{l|\key{if}\ e\ \{ s_1\}\ \key{else}\ \{s_2\};s\},q)\\
\to \text{ob}(o,p,a,\{l|s_2;s\},q)}
\end{array}$$
\end{minipage}}
\caption{\label{fig:sem1}The semantics of Timed ABS.}
\end{figure*}

\emph{Transition rules}  
transform state configurations into new configurations, and are given
in Fig.~\ref{fig:sem1}.  We denote by $a$ the substitution which
represents the attributes of an object and by $l$ the substitution
which represents the local variable bindings of a process.  In the
semantics, different assignment rules are defined for side effect free
expressions (\arulename{Assign1} and \arulename{Assign2}), object
creation (\arulename{New-Object}), method calls
(\arulename{Async-Call}), and future dereferencing
(\arulename{Read-Fut}).  Rule \arulename{Skip} consumes a \key{skip} in
the active process.  Here and in the sequel, the variable $s$ will
match any (possibly empty) statement list. We denote by \textit{idle}
a process with an empty statement list.  
Rules \arulename{Assign1} and \arulename{Assign2} assign the value of
expression $e$ to a variable $x$ in the local variables $l$ or in the
fields $a$, respectively.  Rules \arulename{Cond1} and \arulename{Cond2}
cover the two cases of conditional statements in the same way. (We
omit the rule for \key{while}-loops which unfolds into the conditional.)

\emph{Scheduling.}  Two operations manipulate a process pool $q$;
$pr\append q$ adds a process $pr$ to $q$ and $q\setminus pr$ removes
$pr$ from $q$.  If $q$ is a pool of processes, $\sigma$ a
substitution, $t$ a time value, and $cn$ a configuration, we denote by
\emph{ready}$(q,\sigma,cn)$ the subset of processes from $q$ which are
ready to execute (in the sense that the processes will not directly
suspend or block the object).

Scheduling is captured by the rule \arulename{Schedule}, which applies
when the active process is \emph{idle} and schedules a new process
for execution if there are ready processes in the process pool $q$.
We utilize a scheduling policy in an object
$ob(o,p,\sigma,\textit{idle},q)$, $p$ is an expression representing
the user-defined scheduling policy. This policy selects the process to
be scheduled among the ready processes of the pool $q$.

In order to apply the scheduling policy $p$, which is defined for the
datatype \TYPE{Process} in Timed ABS, to the runtime representation $q$
of the process pool, we lift the processes in $q$ to values of type
\TYPE{Process}.  Let the function \emph{liftall} recursively
transform a pool $q$ of processes to a value of type
\TYPE{List}$\langle$\TYPE{Process}$\rangle$ by repeatedly applying
\emph{lift} to the processes in $q$.  The process identifier of the
scheduled process is used to \emph{select} the runtime representation
of this process from $q$.

Note that in order to evaluate guards on futures, the configuration
$\textit{cn}$ is passed to the \emph{ready} function.  This
explains the use of brackets in the rules, which ensures that
$\nt{cn}$ is bound to the rest of the global system configuration.
The same approach is used to evaluate guards in the rules
\arulename{Await1} and \arulename{Await2} below.

Rule \arulename{Suspend} suspends the active process to the process pool,
leaving the active process \emph{idle}.  Rule \arulename{Await1}
consumes the \key{await}~$g$ statement if $g$ evaluates to true in the
current state of the object, rule \arulename{Await2} adds a
\text{suspend} statement in order to suspend the process if the guard
evaluates to false. 

In rule \arulename{Activation} the function
$\text{bind}(m,o,\bar{v},f,d,c,b,t)$ binds a method call 
to object $o$ in the class of $o$. This results in
 a new process $\{l|s\}$
which is placed in the queue, where 
$l(\texttt{destiny})=f$,
$l(\texttt{method})=m$,
$l(\texttt{arrival})=t$,
$l(\texttt{cost})=c$,
$l(\texttt{deadline})=d$,
$l(\texttt{start})=0$,
$l(\texttt{finish})=0$,
$l(\texttt{crit})=b$,
$l(\texttt{value})=0$,
and where the formal parameters of $m$ are bound to $\many{v}$.

\emph{Durations.}  A statement \key{duration}$(e_1,e_2)$ is reduced to
the runtime statement \key{duration2}$(d_1,d_2)$, in which the
expressions $e_1$ and $e_2$ have been reduced to duration values.
This statement blocks execution on the object until the best case
execution time has passed; i.e., until at least the duration $d_1$
has passed. Remark that time cannot pass beyond duration $d_2$ before
the statement has been executed (see below).

\emph{Method Calls.}  Rule \textsc{Async-Call} sends an invocation
message to $\eval{e}{a\circ l}$ with the unique identity $f$ of a new
future (since $\text{fresh}(f)$), the method name $m$, and 
parameter values $\many{v}$. The identifier of the new future is placed in
the configuration, and is bound to a return value in
\arulename{Return}. The annotations are used to provide a deadline and
a criticality which are passed to the callee with the invocation
message. (The global clock provides a time stamp for the call.)  Rule
\arulename{Return} places the evaluated return expression in the future
associated with the \emph{destiny} variable of the process, and ends
execution after recording the time of process completion in the
\texttt{finish} variable.  Rule \textsc{Read-Fut} dereferences the
future $\text{fut}(f,v)$.  Note that if the future lacks a return
value, the reduction in this object is \emph{blocked}.

\emph{Object creation.}  Rule \textsc{New-Object} creates a new object
with a unique identifier $o'$.  The object's fields are given default
values by $\textit{atts}(C,\eval{\many{e}}{a\circ l},o',c)$, extended
with the actual values $\many{e}$ for the class parameters (evaluated
in the context of the creating process) and $o'$ for \key{this}.  In
order to instantiate the remaining attributes, the process $pr$ is
active (we assume that this process reduces to \emph{idle} if
$\textit{init}(C)$ is unspecified in the class definition, and that it
asynchronously calls \texttt{run} if the latter is specified). The
object gets the scheduler in the annotation $an$ (which is copied from
the class or system default if a scheduler annotation is not
provided).

\emph{Time advance.}  Rule \arulename{Tick} specifies how time can
advance in the system.  We adapt the approach of Real-Time Maude
to Timed ABS and specify a global
time which advances uniformly throughout the global configuration
$cn$, combined with two auxiliary functions: $\textit{adv}(cn,d)$
specifies how the advance of time with a duration $d$ affects
different parts of the configuration $cn$, and $\textit{mte}(cn)$
defines the maximum amount that global time can advance.  At any time,
the system can advance by a duration $d\leq \textit{mte}(cn)$.
However, we are not interested in advancing time by a duration $0$,
which would leave the system in the same state.

The auxiliary functions $\textit{adv}$ and $\textit{mte}$ are defined
in Fig.~\ref{fig:sem2}.  Both have the whole configuration as input
but consider mainly objects since these exhibit time-dependent
behavior.  The function $\textit{mte}$ calculates the maximum time
increment such that no ``interesting'' occurrence (i.e., worst-case
duration expires, duration guard passes) will be missed in any object.
Observe that for statements which are not time-dependent, the maximum
time elapse is $0$ if the statement is enabled, since these statements
are instantaneous, and infinite if not enabled, since time may pass
when the object is blocked.  Hence, $\textit{mte}$ returns the minimum
time increment that lets an object become ``unstuck'', either by
letting its active process continue or enabling one of its suspended
processes.  The function $\textit{adv}$ updates the active and
suspended processes of all objects, decrementing all \texttt{deadline}
values as well as the values in \key{duration} statements and
\key{duration} guards at the head of the statement list in processes.

\begin{figure}[t]
$$\begin{array}{l}
mte(cn_1\mbox{ } cn_2) = min(mte(cn_1), mte(cn_2))\\
 mte(\text{ob}(o, p, a, pr, q)) = \left\{ \begin{array}{ll}
                                          mte(pr) & \mbox{if $pr \neq \text{idle}$}\\
                                          mte(q) & \mbox{if $pr = \text{idle}$}
                                   \end{array}
                              \right. \\
mte(q_1, q_2) = min(mte(q_1), mte(q_2))\\
mte(\{l|s\}) = \left\{ \begin{array}{ll}
                              w & \mbox{if $s = \key{duration2}(b,w);s_2$}\\
                              mte(g) & \mbox{if $s = \key{await}\mbox{ }g;s_2$}\\
                              0 & \mbox{if $s$ is enabled}\\
                              \infty & \mbox{otherwise}
                          \end{array}
                  \right.\\
mte(g) =  \left\{ \begin{array}{ll}
                      max(mte(g_1),mte(g_2)) & \mbox{if }g = g_1 \land g_2\\ 
                      w & \mbox{if }g = \key{duration}(b,w)\\
                      0 & \mbox{if $g$ evaluates to true}\\
                      \infty & \mbox{otherwise}
                      \end{array}
              \right.\\\\
adv(cn_1\mbox{ } cn_2, d) = adv(c_1, d)\mbox{ } adv(c_2, d)\\
adv(\text{ob}(o, p, a, pr, q), d) = \text{ob}(o, p, a, adv(pr, d), adv(q,d))\\
adv((q_1, q_2), d) = adv(q_1, d), adv(q_2,d)\\
adv(\{ l | s\} ,d) = \{ l [ \texttt{deadline} \mapsto l(\texttt{deadline}) -d ] | adv(s,d) \}\\
adv(s,d) = \left\{ \begin{array}{ll}
                      \key{duration2}(b-d, w-d) &\mbox{if }s = \key{duration2}(b,w)\\
                      \key{await}\;adv(g,d) &\mbox{if}\;s =
                      \key{await}\;g \\
                      adv(s_1,d) & \text{if}\; s=s_1;s_2\\
                      s & \text{otherwise}
                      \end{array}
              \right.\\
adv(g,d)=  \left\{ \begin{array}{ll}
                      adv(g_1,d) \land adv(g_2,d) & \mbox{if }g=g_1\land g_2\\
                      \key{duration}(b-d, w-d) &\mbox{if}\;g =
                      \key{duration}(b,w) \\
                       g & \text{otherwise}
                      \end{array}
              \right.
\end{array}$$
\caption{\label{fig:sem2}Functions controlling the advancement of
  time.  Trivial cases for terms \textit{msg}, \textit{fut} have been
  omitted.}
\end{figure}

\end{document}